\documentclass[12pt]{iopart}
\usepackage{fullpage}
\usepackage{cite}
\usepackage{amsfonts}
\usepackage{amssymb}
\usepackage{graphics}
\usepackage{graphicx, color}
\usepackage{epsfig}
\usepackage{color}
\usepackage{ulem}

\newcommand{\be}{\begin{equation}}
\newcommand{\ee}{\end{equation}}
\newcommand{\ba}{\begin{array}}
\newcommand{\ea}{\end{array}}
\newcommand{\ei}{\end{itemize}}
\newcommand{\bea}{\begin{eqnarray}}
\newcommand{\eea}{\end{eqnarray}}
\newcommand{\ben}{\begin{enumerate}}
\newcommand{\een}{\end{enumerate}}
\newcommand{\f}{\frac}
\newcommand{\la}{\langle}
\newcommand{\ra}{\rangle}
\newcommand{\ptl}{\partial}
\newcommand{\al}{\alpha}
\newcommand{\gm}{\gamma}
\newcommand{\Gm}{\Gamma}
\newcommand{\dg}{\dagger}
\newcommand{\nn}{\nonumber}
\newcommand{\pr}{\prime}

\newcommand{\Sxi}{\left[{S}^{x} \right]}
\newcommand{\Syi}{ \left[{S}^{y}\right]}
\newcommand{\Szi}{ \left[{S}^{z} \right]}
\definecolor{brown}{RGB}{255,128,0}
\newcommand{\lx}{ {\color{blue}  \sout{$\ \ \ \blacksquare \ \ $ }}}
\newcommand{\ly}{{ \color{red}  \sout{ $\ \ \blacktriangle \ \ $}} }
\newcommand{\lz}{{ \color{brown}  \sout{ $\ \ \circ \ \ $ }} }
\newcommand{\lp}{{ \color{red}  \sout{ $\ \ * \ \ $ }} }

\begin{document}
\title{Exact Real Time Dynamics of Quantum Spin Systems Using the Positive-P Representation}
\author{R. Ng and E. S. S{\o}rensen}
\address{Department of Physics and Astronomy, McMaster University, 1280 Main St W, L8S4LM, Hamilton, ON, Canada}
\ead{ngry@mcmaster.ca}

\begin{abstract}
We discuss a scheme for simulating the {\it real time} quantum quench dynamics
of interacting quantum spin systems within the positive-P formalism. As model
systems we study the transverse field Ising model as well as the Heisenberg
model undergoing a quench away from the classical ferromagnetic ordered state and antiferromagnetic N\'{e}el state, depending on the sign of the Heisenberg exchange interaction.
The connection to the positive-P formalism as it is used in quantum optics is
established by  mapping the spin operators on to Schwinger bosons. In doing so,
the dynamics of the interacting quantum spin system is mapped onto a set of Ito
stochastic differential equations (SDEs) the number of which scales {\it
linearly} with the number of spins, $N$, compared to an exact solution
through diagonalization that in the case of the Heisenberg model would
require matrices exponentially large in $N$. This mapping is {\it exact} and
can be extended to higher dimensional interacting systems as well as to
systems with an explicit coupling to the environment.
\end{abstract}
\submitto{\JPA}
\maketitle

\section{Introduction}
\label{ sn: Introduction}

The {\it real time} quantum dynamics following a
quench~\cite{CC1,CC2,Kollath07,Manmana07,Barmettler09,Roux09,Langer09,Roux10,quantumquenches, 3rdquant, bethe, bethe2, MERA, Fioretto, Fagotti} is a problem of considerable current
interest. Here, our focus is on methods applicable to this problem that are in
principle exact (up to controllable errors) and we leave approximate methods
aside. Unfortunately, standard quantum Monte Carlo techniques yield results in
the imaginary time domain and requires an explicit analytic continuation to
access real times, a notoriously difficult procedure. For lattice based models
it is possible to perform exact diagonalization but for a $N$ site quantum spin
system the size of the Hilbert space is exponential in $N$, severely limiting
the applicability of this method. In recent years methods rooted in the density
matrix renormalization group (DMRG) such as TEBD~\cite{TEBD} and
t-DMRG~\cite{tDMRG} have been developed to study real-time dynamics of
one-dimensional systems. Most recently the infinite size TEBD (iTEBD) has been
tuned to yield results for the time dependence of the transverse field Ising
model (TFIM) out to relatively large times of order
$tJ/\hbar\sim6-10$~\cite{Banuls} as well as in the $XXZ$ and related spin chain models $tJ/\hbar\sim20$~\cite{Barmettler09,quantumquenches} and often
times scales of order $tJ/\hbar\sim 100$ can be accessed~\cite{Gobert05}. How well such methods will perform in higher
dimensions or in the presence of a coupling to the environment is presently
a point of intense research and very promising progress have been made~\cite{Verstraete04,Zwolak04,Hartmann09,Clark10}. Here we  investigate an alternative approach for studying the dynamics
of interacting quantum spin systems using quantum phase space methods, in
particular, the positive-P representation (PPR)~\cite{positive} of the density
operator. As model systems we have studied the one-dimensional transverse field
Ising model (TFIM) as well as the Heisenberg model. This approach is quite
general and can be extended to higher dimensional interacting quantum spin
systems and to open systems with an explicit coupling to the environment.

In general, quantum phase space methods map the dynamics of bosonic operators onto the stochastic evolution of complex phase space variables~\cite{positive}. Using the positive-P representation, we can easily calculate the expectation values of any normal-ordered products of creation and annihilation operators by calculating the stochastic averages of their equivalent representation in terms of phase-space variables.
This is carried out in two steps, first we use Schwinger bosons to replace the Heisenberg spin operators and then employ the positive-P representation. The PPR converts the master equation into a Fokker-Planck equation (FPE) which can then be mapped onto a set of coupled, complex Ito  stochastic differential equations (SDEs). The number of SDEs to simulate scales linearly with the number of spins in the system, $N$, in contrast to an exact diagonalization approach. 

To illustrate the feasibility of this approach we study the dynamics of the transverse-field Ising model (TFIM)
as well as the isotropic ferromagnetic (FM) Heisenberg model subject to a quantum quench at $T=0$. The different models are related through the anisotropy parameter, $\Delta/J$. The spin chains are prepared in the ferromagnetic state at $t=0$  whenever we assume a ferromagnetic Heisenberg model, and evolved by including the transverse magnetic field term at $t\geq0$. We calculate the time evolution of the expectation values of the spin operators: $ \Sxi, \Syi \Szi$, which is an average of the individual components over the entire lattice. The averaging is allowed because of the translational symmetry of the system. In addition, we also calculate the results of $\hat{S}_{z}$ nearest neighbour correlation functions: $\left[ \hat{S}_{i}^{z} \hat{S}_{i+1}^{z} \right]$ for the TFIM. In order to verify the validity of our results, we in all cases compare them with results from exact diagonalization obtaining excellent agreement.

In a bid to fully take advantage of the PPR, we also attempt to explore finite size effects by simulating lattices sizes of up to $100$ spins for the FM isotropic model and $10$ spins for the antiferromagnetic (AFM) anisotropic model. Finite size effects are more noticeable in the AFM Hamiltonian and for the latter, the natural choice for an initial state is the classical N\'{e}el state. 

Since the PPR is well-established in quantum optics, we will relegate the details of the formalism to ~\ref{sn: positiveP}.  Readers who are already familiar with the PPR may continue to section~\ref{sn: Schwinger Bosons} where Schwinger bosons are employed to map the spin operators onto bosonic operators. The resulting SDEs are derived in this section with more explicit details layed out in~\ref{sn: additional details}. In section~\ref{sn: results and discussion}, the results of the TFIM $(\Delta/J =0.0)$ and the isotropic $(\Delta/J=1.0 )$ Heisenberg model are compared with exact diagonalization calculations. We also carry out a brief discussion on the possbility of extending simulation life times by potentially using the gauge-P representation~\cite{gaugeP} instead. In section~\ref{sn: finite size effects}, we present our results for finite size effects in both the anisotropic AFM and the isotropic FM Hamiltonian and discuss our findings. Results and a short discussion on the correlation functions can be found in section~\ref{ssn: corr functions}. The conclusion is presented  in section~\ref{sn: Conclusion}. 

\section{Using Schwinger Bosons to derive SDEs}
\label{sn: Schwinger Bosons}
The PPR is based on bosonic coherent states and is only directly applicable to Hamiltonians written in terms of
bosonic annihilation and creation operators.  In order to apply it to the Heisenberg model or any spin Hamiltonian,
we therefore need to rewite the spin operators in terms of bosonic operators. A convenient way of doing this
is by employing the Schwinger boson representation~\cite{Schwinger, Sakurai} and we will demonstrate how
it can be applied to the Heisenberg model. A similar approach, based on Schwinger bosons, was previously
applied to the study of spontaneous emission non-interacting two-level atoms~\cite{SpontEmission} in quantum optics.

The Heisenberg Hamiltonian with FM ($J>0$) or AFM interaction ($J<0$) subject to a quench in the $x$-direction at $t\geq0$ is given by:

\be
\label{eq: heisenberg}
\hat{H} = -J \sum_{\la i,j \ra} {\bf \hat{S}}_{i}\cdot { \bf \hat{S} }_{j} - h(t) \sum_{i} \hat{S}^{x}_{i}, \ \ \ h(t) = \left\{ \ba{c} h, \ t\geq 0 \\                                                                                                                          0, \ t<0\ea \right.
\ee
and can be written in terms of the usual raising and lowering operators, ${\bf \hat{S} }^{\pm} = \mathbf{\hat{S}^{x}} \pm i \mathbf{\hat{S}^{y}}$. If we allow anisotropy in the transverse direction\footnote{The transverse direction is relative to the quantization axis which we have taken to be the $z$-axis.}, then the Hamiltonian takes the following form

\be
\label{eq: heisenberg with anisotropy}
\hat{H}_\mathrm{Heis} = -\sum_{\la i,j \ra} \left[ J \hat{S}^{z}_{i} \hat{S}^{z}_{j}  + \Delta\f{1}{2} ( \hat{S}_{i}^{+} \hat{S}_{j}^{-} + \hat{S}_{i}^{-} \hat{S}_{j}^{+} ) \right] - \f{1}{2}h(t) \sum_{i} \left[\hat{S}^{+}_{i} + \hat{S}^{-}_{i} \right]
\ee
where $\la i,j \ra$ indicates nearest-neighbor pairs and $\Delta/J$ is a measure of anisotropy.  The two models which we first examined were the (i) TFIM $(\Delta/J=0)$:

\be
\label{eq: TFIM}
\hat{H}_\mathrm{TFIM} = -\sum_{\la i,j \ra} \left[ J \hat{S}^{z}_{i} \hat{S}^{z}_{j} \right] - \f{1}{2}h(t) \sum_{i} \left[\hat{S}^{+}_{i} + \hat{S}^{-}_{i} \right]
\ee
and the (ii) isotropic Heisenberg model (see Eq.~\ref{eq: heisenberg with anisotropy}) with an anisotropy of $\Delta/J=1.0$.

 The Schwinger boson representation of spins (setting $\hbar = 1$) is given by:

\be
\label{eq: schwinger}
\hat{S}^{+} \rightarrow  \hat{b} \hat{a}^{\dg}, \hat{S}^{-} \rightarrow  \hat{b}^{\dg} \hat{a}, \hat{S}^{z} \rightarrow \f{1}{2} \left( \hat{a}^{\dg}\hat{a} - \hat{b}^{\dg}\hat{b}\right).
\ee
where $\hat{a}$ and $\hat{b}$ represent two types of bosons and the following commutation relations:

\bea
\left[ \hat{S}^{+}, \hat{S}^{-} \right] \rightarrow & \left[ \hat{a}^{\dg} \hat{b}, \hat{b}^{\dg} \hat{a}  \right] &  =  \hat{a}^{\dg} \hat{a} - \hat{b}^{\dg} \hat{b} \rightarrow 2\hat{S}^{z}, \nn\\
\left[ \hat{S}^{+}, \hat{S}^{z} \right] \rightarrow & \left[ \hat{a}^{\dg} \hat{b} , \f{1}{2} ( \hat{a}^{\dg} \hat{a} - \hat{b}^{\dg} \hat{b} ) \right] & = -\hat{a}^{\dg} \hat{b} \rightarrow - \hat{S}^{+}, \nn \\
\left[ \hat{S}^{-}, \hat{S}^{z} \right] \rightarrow & \left[ \hat{b}^{\dg} \hat{a} , \f{1}{2} ( \hat{a}^{\dg} \hat{a} - \hat{b}^{\dg} \hat{b} ) \right] &  = \hat{b}^{\dg} \hat{a} \rightarrow \hat{S}^{-}
\eea
demonstrate that the commutation relations of the spin operators are indeed preserved. This is a necessary requirement for a successful mapping. With the Schwinger representation, the two states of a spin-$1/2$ particle are now described by either an $\hat{a}-$boson or a $\hat{b}$-boson per site. A spin-up state: $| \uparrow \ra$ is the same as having a single $\hat{a}$-boson whereas a spin down-state: $| \downarrow \ra$ is the same as having a single $\hat{b}$-boson. We can therefore replace the spin operators in Eq.~\ref{eq: heisenberg with anisotropy} and Eq.~\ref{eq: TFIM}  with the bosonic mapping in Eq.~\ref{eq: schwinger} without altering the physics.

 As the PPR is well-established\footnote{See~\cite{SpontEmission, nonlinDamping, secondharmonic, dynamicalnoise, gaugeI, 150000} for successful applications of the PPR.}, we will relegate a brief review of the formalism to  ~\ref{sn: positiveP} . Additional technical details pertaining to the specific examples in this paper can be found in ~\ref{sn: additional details}.  For brevity we will present the  derivations for only the TFIM (see Eq.~\ref{eq: TFIM}) where $\Delta/J=0$.

 Using Eq.~\ref{eq: schwinger}, the equivalent bosonic Hamiltonian for the TFIM is given by

\be
\label{eq: schwinger hamiltonian}
\fl \hat{H} = -\f{J}{4} \sum_{\la i,j\ra} \left( \hat{a}^{\dg}_{i} \hat{a}_{i} \hat{a}^{\dg}_{j} \hat{a}_{j}  - \hat{a}^{\dg}_{i} \hat{a}_{i}\hat{b}^{\dg}_{j} \hat{b}_{j} - \hat{b}^{\dg}_{i} \hat{b}_{i} \hat{a}^{\dg}_{j} \hat{a}_{j} + \hat{b}^{\dg}_{i} \hat{b}_{i}\hat{b}^{\dg}_{j} \hat{b}_{j}\right) -  \left(h(t) \sum_{i} \hat{a}_{i}^{\dg} \hat{b}_{i} + \hat{b}^{\dg}_{i} \hat{a}_{i} \right).
\ee

Now if we take our system to be closed, its dynamics can be captured via the master equation for the density operator, i.e.

\be
\label{eq: master equation}
\f{d}{dt} \hat{\rho} = -\f{i}{\hbar} \left[ \hat{H}, \hat{\rho} \right],
\ee
which allows us to use a generalized prescription of the PPR. In principle, it is also possible to calculate open system dynamics by including a Liouvillan term in Eq.~\ref{eq: master equation}: $\hat{L}[\hat{\rho}]$\footnote{The liouvillian term models the effect of the environment on the system}, and so this approach is by no means limited to closed system.

 To proceed, we first write our density operator in terms of  a direct product of projection operators for each site,  i.e.

\be
\hat{\Lambda}(\vec{\al}, \vec{\al}^{+}, \vec{\beta}, \vec{\beta}^{+}) = \prod_{i=0}^{N-1} \otimes \f{| \al_{i} \ra \la \al^{+\ast}_{i} |}{ \la \al_{i}^{+\ast} | \al_{i} \ra }\otimes \f{| \beta{i} \ra \la \beta^{+\ast}_{i} |}{ \la \beta_{i}^{+\ast} | \beta_{i} \ra }
\ee
where $\vec{\al} = ( \al_{0}, \dots, \al_{N-1})$,  $\vec{\al}^{+} = ( \al^{+}_{0}, \dots, \al^{+}_{N-1})$, $\vec{\beta} = ( \beta_{0}, \dots, \beta_{N-1})$
and
$\vec{\beta}^{+} = ( \beta^{+}_{0}, \dots, \beta^{+}_{N-1})$
so that

\be
\hat{\rho} = \int  P(\vec{\al}, \vec{\al}^{+}, \vec{\beta}, \vec{\beta}^{+}) \hat{\Lambda}(\vec{\al}, \vec{\al}^{+}, \vec{\beta}, \vec{\beta}^{+}) d^{2} \vec{\al} d^{2} \vec{\al}^{+}d^{2} \vec{\beta} d^{2} \vec{\beta}^{+}.
\ee
We can then use the usual correspondence relations (see Eq.~\ref{eq: correspondence relations}) to obtain an FPE (see Eq.~\ref{eq: FPE}) for the PPR distribution function: $P(\vec{\al}, \vec{\al}^{+}, \vec{\beta}, \vec{\beta}^{+)}$. A particular factorization of the diffusion matrix results in a noise matrix which gives us a set of Ito stochastic differential equations for $4N$ of our phase space variables, i.e.

\bea
\label{eq: SDE al}
\fl d \al_{i}    =    \left\{ \f{i J }{4 \hbar} \al_{i}  \left[ (n_{i+1}^{\al} - n_{i+1}^{\beta} ) + (n^{\al}_{i-1} - n^{\beta}_{i-1}  )\right] + \f{ih(t)}{2\hbar} \beta_{i} \right\}dt  \nn \\
\fl    +\f{1}{2}\sqrt{\f{iJ}{2\hbar}} \left[ \sqrt{\al_{i} \al_{i+1}} ( dW_{2i}^{\al} + i dW_{2i+1}^{\al} ) - \sqrt{\al_{i}  \al_{i-1}} ( dW_{2i-2}^{\al} - i dW^{\al}_{2i-1})\right] \nn \\
\fl +\f{i}{2}\sqrt{\f{iJ}{2\hbar}} \left[ \sqrt{\al_{i} \beta_{i-1}} ( dW_{2i-2}^{\al\beta} + i dW_{2i-1}^{\al \beta} ) - \sqrt{\al_{i} \beta_{i+1}} ( dW_{2i}^{\beta \al} - i dW^{\beta \al}_{2i+1})\right]
\eea

\bea
\label{eq: SDE beta}
\fl  d \beta_{i}  =    \left\{ \f{i J }{4 \hbar} \beta_{i} \left[ (n_{i+1}^{\beta} - n_{i+1}^{\al} ) + (n^{\beta}_{i-1} - n^{\al}_{i-1}  )\right] + \f{ih(t)}{2\hbar} \al_{i} \right\}dt \nn \\
\fl   +\f{1}{2}\sqrt{\f{iJ}{2\hbar}} \left[ -\sqrt{\beta_{i} \beta_{i+1}} ( dW_{2i}^{\beta} + i dW_{2i+1}^{\beta} ) - \sqrt{\beta_{i} \beta_{i-1}} ( dW_{2i-2}^{\beta} - i dW^{\beta}_{2i-1})\right] \nn \\
\fl  +\f{i}{2}\sqrt{\f{iJ}{2\hbar}} \left[ -\sqrt{\al_{i+1} \beta_{i}} ( dW_{2i}^{\al\beta} - i dW_{2i+1}^{\al \beta} ) - \sqrt{\al_{i-1} \beta_{i}} ( dW_{2i-2}^{\beta \al} + i dW^{\beta \al}_{2i-1}) \right]
\eea

\bea
\label{eq: SDE al+}
\fl d \al^{+}_{i}   =  \left\{ \f{i J }{4 \hbar} \al^{+}_{i} \left[ (n_{i+1}^{\beta} - n_{i+1}^{\al} ) + (n^{\beta}_{i-1} - n^{\al}_{i-1}  )\right] - \f{ih(t)}{2\hbar} \beta^{+}_{i} \right\} dt  \nn \\
\fl +\f{i}{2}\sqrt{\f{iJ}{2\hbar}} \left[ -\sqrt{\al^{+}_{i} \al^{+}_{i+1}} ( dW_{2i}^{\al^{+}} + i dW_{2i+1}^{\al^{+}} ) - \sqrt{\al^{+}_{i} \al^{+}_{i-1}} ( dW_{2i-2}^{\al^{+}} - i dW^{\al^{+}}_{2i-1})\right] \nn \\
\fl +\f{1}{2}\sqrt{\f{iJ}{2\hbar}} \left[ -\sqrt{\al^{+}_{i} \beta^{+}_{i-1}} ( dW_{2i-2}^{\al^{+}\beta^{+}} + i dW_{2i-1}^{\al^{+} \beta^{+}} ) - \sqrt{\al_{i}^{+} \beta^{+}_{i+1}} ( dW_{2i}^{\beta^{+} \al^{+}} - i dW^{\beta^{+} \al^{+}}_{2i+1})\right]
\eea

\bea
\label{eq: SDE beta+}
\fl d \beta^{+}_{i}   =  \left\{ \f{i J }{4 \hbar} \beta^{+}_{i} \left[ (n_{i+1}^{\al} - n_{i+1}^{\beta} ) + (n^{\al}_{i-1} - n^{\beta}_{i-1}  )\right]  -\f{ih(t)}{2\hbar} \al_{i}^{+} \right\}dt \nn \\
\fl +\f{i}{2}\sqrt{\f{iJ}{2\hbar}} \left[ -\sqrt{\beta^{+}_{i} \beta^{+}_{i+1}} ( dW_{2i}^{\beta^{+}} + i dW_{2i+1}^{\beta^{+}} ) - \sqrt{\beta^{+}_{i} \beta^{+}_{i-1}} ( dW_{2i-2}^{\beta^{+}} - i dW^{\beta^{+}}_{2i-1})\right] \nn \\
\fl +\f{1}{2}\sqrt{\f{iJ}{2\hbar}} \left[ -\sqrt{\al^{+}_{i+1} \beta^{+}_{i}} ( dW_{2i}^{\al^{+}\beta^{+}} - i dW_{2i+1}^{\al^{+} \beta^{+}} ) - \sqrt{\al^{+}_{i-1} \beta^{+}_{i}} ( dW_{2i-2}^{\beta^{+} \al^{+}} + i dW^{\beta^{+} \al^{+}}_{2i-1})\right],
\eea
where $i=0\dots N-1$ labels the vector components and we have defined $n_{i}^{\al} = \al_{i}^{+}\al_{i}$ and $n_{i}^{\beta} =  \beta_{i}^{+}\beta_{i}$, which are complex phase space functions representing the number of $\hat{a}$ and $\hat{b}$-bosons (per site $i$) respectively. With this particular choice of noise matrix, we have introduced eight $2N\times1$ Wiener increment vectors with the usual statistical properties that $\la dW^{x}_{i} dW^{y}_{j} \ra = dt \delta_{xy} \delta_{ij}$  and $\la dW^{x}_{i}\ra =0$, where $i=0 \dots N-1$ and $x,y =\al, \al^{+}, \beta, \beta^{+}, \beta\al, \al\beta, \beta^{+} \al^{+}, \al^{+}\beta^{+}$ labels each Wiener increment vector. We would like to point out that the subscript labels of the Wiener increment vector are not unique and the labeling scheme\footnote{Note that with the inclusion of periodic boundary conditions: $\al_{-1} \rightarrow \al_{N-1}$ and $\al_{N} \rightarrow \al_{0}$. However since there are $2N\times1$ Wiener increments, then it is periodic in $2N$ instead. For e.g. $dW^{x}_{-1} = dW^{x}_{2N-1}$ and $dW^{x}_{2N} = 0$.} was chosen simply for convenience (see~\ref{sn: additional details}).

\subsection{Inclusion of Anisotropy}
 Had we begun with the full anisotropic Hamiltonian in Eq.~\ref{eq: heisenberg with anisotropy} instead and carried out the same steps as in section~\ref{sn: Schwinger Bosons}, it can be shown that anisotropy is included by adding the following expressions into the drift terms of Eq.~\ref{eq: SDE al} to Eq.~\ref{eq: SDE beta+}:

\bea
\label{eq: modify drift}
d \al_{i} & \sim & +\f{i \Delta}{2\hbar} \beta_{i} (m_{i-1} + m_{i+1}) dt   \\
d \beta_{i} & \sim & +\f{i \Delta}{2\hbar} \al_{i} (m^{+}_{i-1} + m^{+}_{i+1}) dt  \\
d \al^{+}_{i} & \sim & -\f{i \Delta}{2\hbar} \beta^{+}_{i} (m^{+}_{i-1} + m^{+}_{i+1}) dt   \\
d \beta^{+}_{i} & \sim & -\f{i \Delta}{2\hbar} \al^{+}_{i} (m^{+}_{i-1} + m^{+}_{i+1}) dt
\eea
where the following shorthand  $m_{i}= \al_{i} \beta_{i}^{+}, m^{+}_{i}=\al_{i}^{+} \beta_{i}$ was used. For the stochastic terms however, only the mixed derivative diffusion terms (i.e. those containing  $\al \beta$ and $\al^{+} \beta^{+}$) are modified in the following way

\bea
\label{eq: modify noise}
\fl  d \al_{i} & \sim & +\f{i}{2}\sqrt{\f{i}{2\hbar}} \left[ -\sqrt{J\al_{i} \beta_{i-1} -2\Delta\beta_{i} \al_{i-1} } ( \dots \dots ) - \sqrt{J\al_{i} \beta_{i+1}- 2\Delta \al_{i+1} \beta_{i} } ( \dots \dots)\right] \\
\fl d \beta_{i} & \sim & +\f{i}{2}\sqrt{\f{i}{2\hbar}} \left[ -\sqrt{J\beta_{i} \al_{i+1} -2\Delta\beta_{i+1} \al_{i} } ( \dots \dots ) - \sqrt{J\beta_{i}\al_{i-1} - 2\Delta \al_{i} \beta_{i-1} } ( \dots \dots)\right]   \\
\fl d \al^{+}_{i} & \sim &  +\f{i}{2}\sqrt{\f{i}{2\hbar}} \left[ -\sqrt{J\beta^{+}_{i-1} \al^{+}_{i} -2\Delta\beta^{+}_{i} \al^{+}_{i-1} } ( \dots \dots ) - \sqrt{J\beta^{+}_{i+1} \al^{+}_{i}- 2\Delta \al^{+}_{i+1} \beta^{+}_{i} } ( \dots \dots)\right] \\
\fl  d \beta^{+}_{i} & \sim & +\f{i}{2}\sqrt{\f{i}{2\hbar}} \left[ -\sqrt{J\beta^{+}_{i} \al^{+}_{i-1} -2\Delta\beta^{+}_{i-1} \al^{+}_{i} } ( \dots \dots ) - \sqrt{J\beta^{+}_{i} \al^{+}_{i+1}- 2\Delta \al^{+}_{i} \beta^{+}_{i+1} } ( \dots \dots)\right]
\eea
where the terms in  $(\dots \dots )$ represent the same Wiener increment combinations as in Eq.~\ref{eq: SDE al} to~\ref{eq: SDE beta+}.  The Ito SDEs we have derived are able to describe other types of spins models such as the  XY model and the XYZ model (to name a few), just by adjusting or including a few parameters. For the last two cases, we would have to take a trivial generalization in the derivations by introducing two different anisotropy terms in Eq.~\ref{eq: heisenberg with anisotropy}. An informative review article on the the quantum quench dynamics of other variants of the Heisenberg Hamiltonian using other numerical methods can be found in~\cite{quantumquenches}.

\section{Results and Discussion}
\label{sn: results and discussion}
 To test our formalism, we first simulated the FM $(J > 0)$ spin Hamiltonian for the TFIM $(\Delta/J=0)$ and the isotropic Heisenberg Hamiltonian $( \Delta/J = 1.0)$ in Eq.~\ref{eq: heisenberg with anisotropy} for high $(h/J=10)$ and low $(h/J=0.5 )$ field values. This was compared to results from exact diagonalization calculations using a small system with $N=4$ spins. The Stratanovich version of the SDES \footnote{The Stratanovich correction terms worked out to be zero and hence the Stratanovich form of the SDEs from Eq.~\ref{eq: SDE al} to Eq.~\ref{eq: SDE beta+} have the exact same form as the derived Ito SDEs.} in Eq.~\ref{eq: SDE al} to Eq.~\ref{eq: SDE beta+}  were simulated using a semi-implicit Stratanovich algorithm as they are known to exhibit superior convergence properties~\cite{Mortimer}.
 To track the dynamics of the system,  we calculated the expectation values of all three spin components at each site $i$:  $\la {S}^{x}_{i} \ra, \la {S}^{y}_{i} \ra, \la {S}^{z}_{i} \ra$. Using the translation symmetry of the system, we further averaged them over the entire lattice to obtain an average expectation value of the spin components per site: $\Sxi, \Syi, \Szi$. These expectation values were calculated using the stochastic averages of their respective phase space functions, i.e.

\begin{figure}
\begin{center}
    \includegraphics{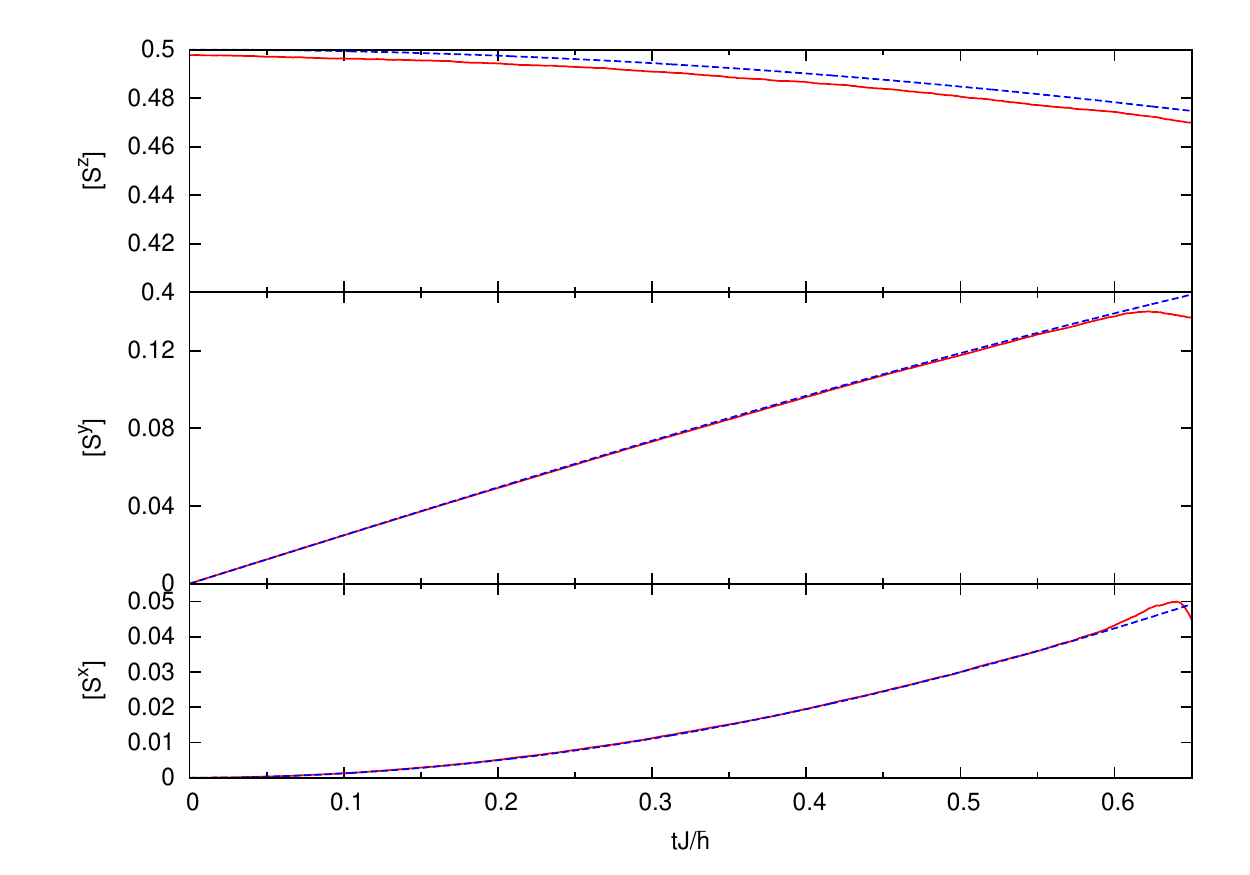}
\end{center}\caption{TFIM following a transverse quench. From top to bottom: plots of $\Sxi, \Syi, \Szi $ vs $t J/ \hbar$ respectively. The stochastic averages, $\la \la . \ra\ra$ are given by red solid lines while exact diagonalization results are represented by the green dashed lines. Simulation parameters: $ N = 4, n_{traj}=10^{6}, dt=0.001, h/J = 0.5 , \Delta/J=0.0$. Agreement remains good till approximately $t J/\hbar = 0.6$.  }\label{fig: fig1}
\end{figure}

\be
\label{eq: obs Sx}
\Sxi =  \sum_{i=0}^{N-1} \la \f{1}{2} ( \hat{a}_{i}^{\dg} \hat{b}_{i} + \hat{b}_{i}^{\dg} \hat{a}_{i} )\ra =  \sum_{i=0}^{N-1} \la \la \f{1}{2} (\al^{+}_{i} \beta_{i} + \beta^{+}_{i}\al_{i} ) \ra \ra,
\ee
\be
\label{eq: obs Sy}
\Syi  =  \sum_{i=0}^{N-1} \la \f{1}{2i} ( \hat{a}_{i}^{\dg} \hat{b}_{i} - \hat{b}_{i}^{\dg} \hat{a}_{i} )\ra =  \sum_{i=0}^{N-1} \la \la \f{1}{2i} (\al^{+}_{i} \beta_{i} - \beta^{+}_{i}\al_{i} ) \ra \ra,
\ee
\be
\label{eq: obs Sz}
\Szi =  \sum_{i=0}^{N-1} \la \f{1}{2} ( \hat{a}_{i}^{\dg} \hat{a}_{i} - \hat{b}_{i}^{\dg} \hat{b}_{i} )\ra =  \sum_{i=0}^{N-1} \la \la \f{1}{2} (\al^{+}_{i} \al_{i} - \beta^{+}_{i}\beta_{i} ) \ra \ra,
\ee
where $\la \la . \ra \ra$ denotes a stochastic average.

\begin{figure}
\begin{center}
    \includegraphics{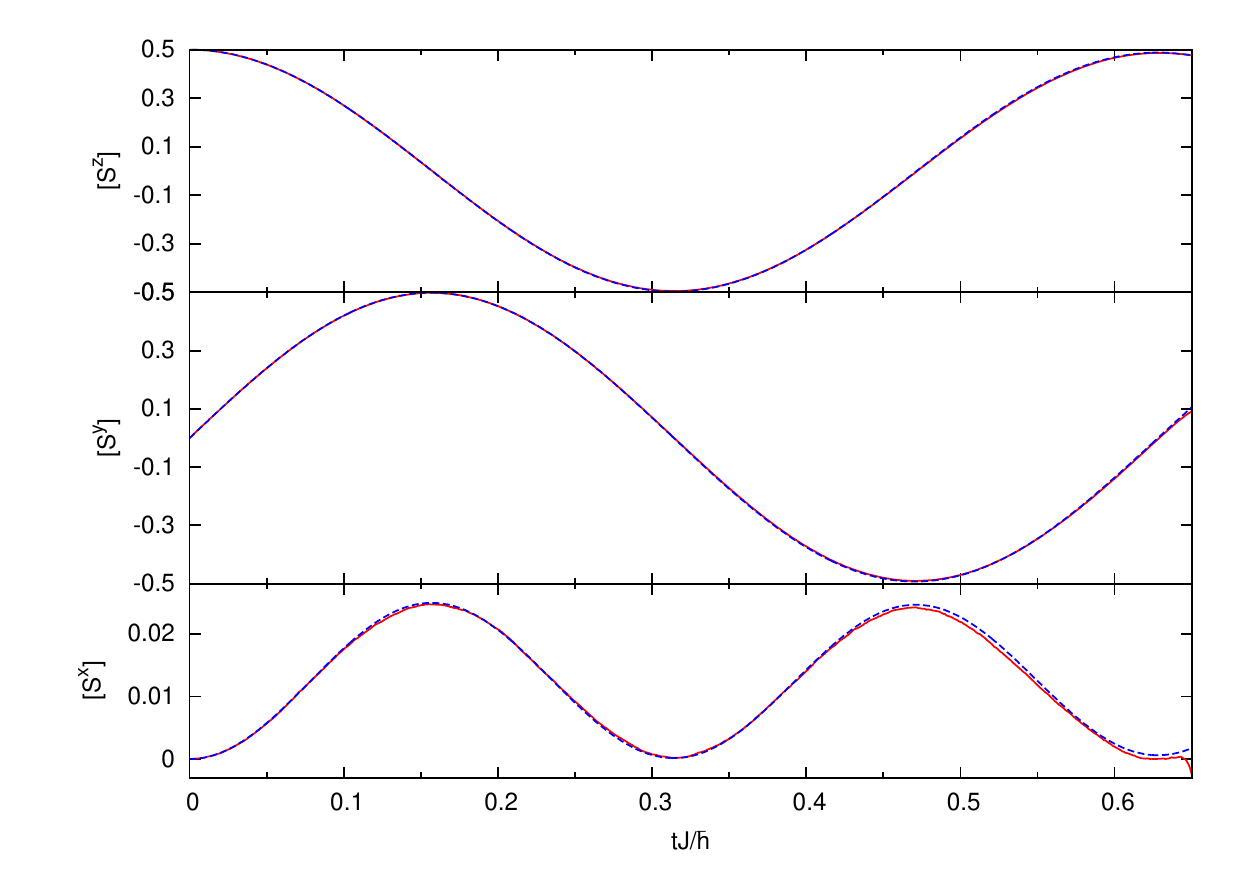} 
\end{center}\caption{TFIM following a transverse quench. From top to bottom: plots of $\Sxi, \Syi, \Szi $ vs $t J/ \hbar$ respectively. The stochastic averages, $\la \la . \ra\ra$ are given by red solid lines while exact diagonalization results are represented by the green dashed lines. Simulation parameters: $ N = 4, n_{traj}=2\times 10^{5}, dt=0.001, h/J = 10.0, \Delta/J=0.0$. Agreement remains good till approximately $t J/\hbar = 0.65$. }\label{fig: fig2}
\end{figure}

\begin{figure}
\begin{center}
    \includegraphics{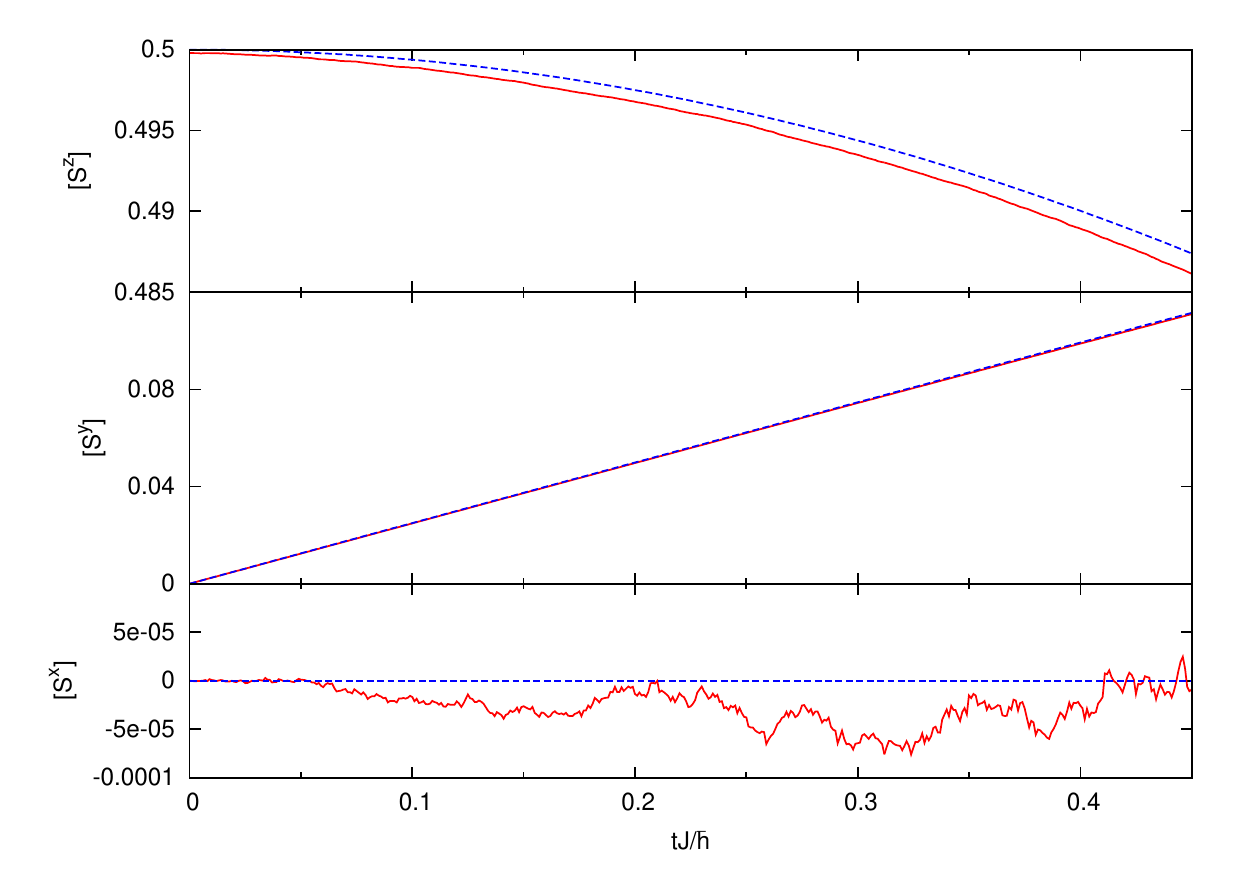} 
\end{center}\caption{Isotropic Heisenberg model following a transverse quench. From top to bottom: plots of $\Sxi, \Syi, \Szi$ vs $tJ/\hbar$ respectively. The stochastic averages, $\la \la . \ra\ra$ are given by red solid lines while exact diagonalization results are represented by green dashed lines. Simulation parameters: $N = 4, n_{traj}= 10^{6}, dt=0.001, h/J = 0.5, \Delta/J=1.0$. Agreement remains good and both results are nearly indistinguishable. The simulations diverge at approximately $tJ/\hbar = 0.45$. }\label{fig: fig3}
\end{figure}

\begin{figure}
\begin{center}
    \includegraphics{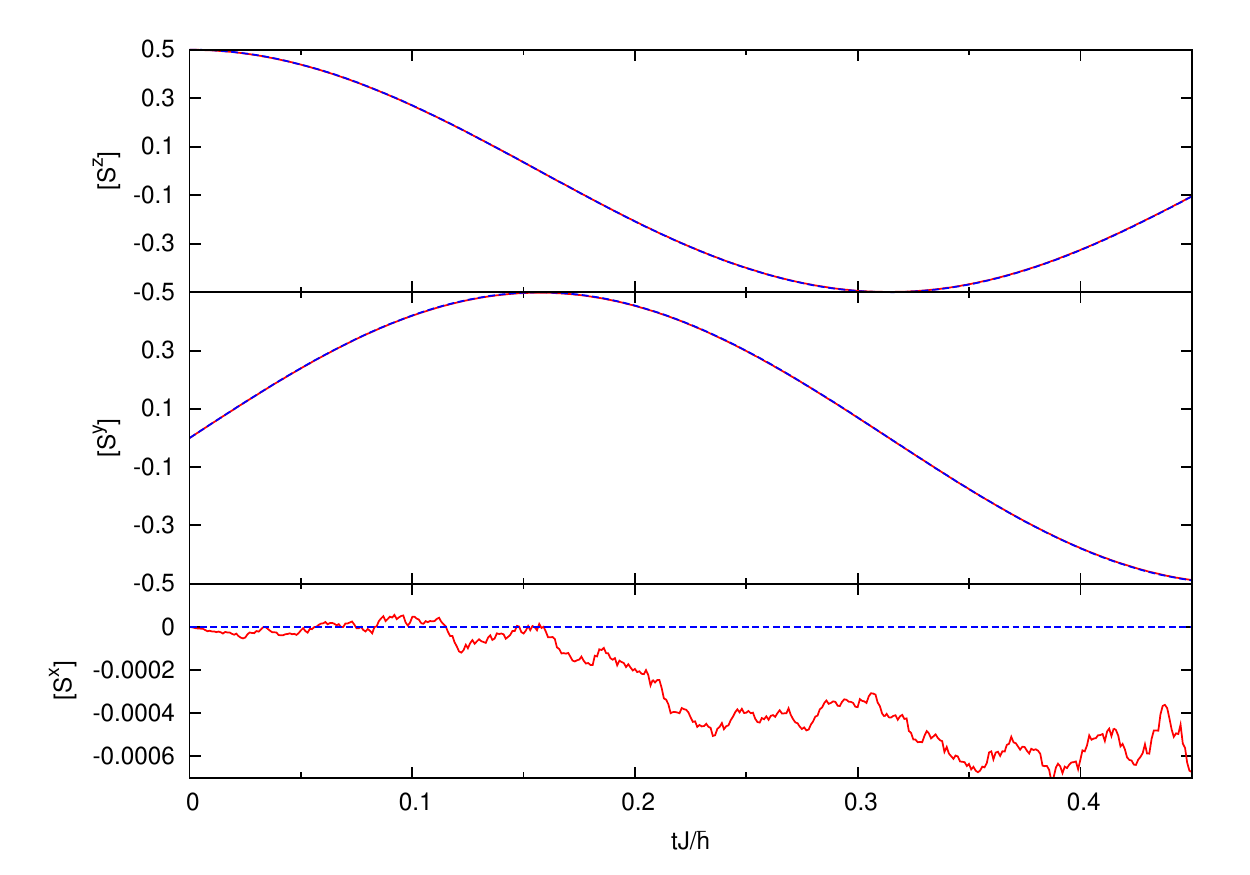}
\end{center}\caption{Isotropic Heisenberg model following a transverse quench. From top to bottom: plots of $\Sxi, \Syi, \Szi$ vs $tJ/\hbar$ respectively. The stochastic averages, $\la \la . \ra\ra$ are given by red solid lines while exact diagonalization results are represented by green dashed lines. Simulation parameters: $N = 4, n_{traj}= 10^{5}, dt=0.001, h/J = 10.0, \Delta/J=1.0$. Agreement remains good and both results are nearly indistinguishable. The simulations diverge at approximately $tJ/\hbar = 0.45$. }\label{fig: fig4}
\end{figure}

The initial state of the system was taken to be the classical ferromagnetic state: $|  \uparrow \uparrow \dots \uparrow \ra$ and the dynamics were observed for $t\geq 0$ during which a transverse field is turned on. The results for the TFIM are shown in Fig.~\ref{fig: fig1} and Fig.~\ref{fig: fig2} for different field strengths while the results for the isotropic $(\Delta/J = 1.0)$ model are shown in Fig.~\ref{fig: fig3} and Fig.~\ref{fig: fig4}. Both models show good agreement with exact diagonalization calculations.

 The only drawback of the PPR is that the simulations are usually valid only for relatively short lifetimes (roughly $tJ/ \hbar\sim 0.45-0.65$ for the models examined) before sampling errors caused by diverging trajectories take over. In Fig.~\ref{fig: fig1} for example, the onset of the effects of diverging trajectories can be seen at around $t J/\hbar \sim 0.58$  where a deviation of the SDE results and exact calculations begin to appear. However, for the time scales where the simulations remain finite, it does yield good results.

One should not be alarmed as this is a common problem associated in using the PPR and can be attributed to the nature of the SDEs derived and not due to a non-converging numerical algorithm~\cite{gilchrist, optimization, gaugeII}.  In fact, Deuar~\cite{thesis} examined this issue when applying the PPR to the exact dynamics of many-body systems.  If we abide by Deuar's findings strictly, we see that there are no drift and noise divergences present in the SDEs in Eq.~\ref{eq: SDE al}-~\ref{eq: SDE beta+}. However, we suspect drift terms of the form $\sim i X_{i} \left[ (\mp n_{i+1}^{\al} \pm n_{i+1}^{\beta} ) + (n^{\al}_{i-1} \pm n^{\beta}_{i-1}  )\right] $, where $X_{i} = \al_{i}, \al^{+}_{i}, \beta, \beta^{+}$ can be problematic. This is because if we take into consideration the translational symmetry of the system, then we can approximately say that
\be
i X_{i} \left[ \left( \mp n^{\al}_{i+1} \pm n^{\beta}_{i+1} \right) + \left( \mp n^{\al}_{i-1} \pm n^{\beta}_{i-1} \right) \right] \approx 2 i X_{i} \left(\mp n^{\al}_{i} \pm n^{\beta}_{i} \right),
\ee
which now clearly exhibits offending terms~\cite{thesis} that cause trajectories to escape to infinity, since $dX_{i} \sim X_{i}^{2} \left[\dots \right] dt  + \dots$.

The gauge-P representation~\cite{thesis, gaugeII, stochasticgauges1, stochasticgauges2, gaugeP} was developed  to specifically deal with such drift instabilities. In the gauge-P representation,  arbitrary gauge functions, $\left\{ g_{k}\right\}$ can be introduced into the SDEs whose effect is a modification of the deterministic evolution. This can be done at the expense of introducing another stochastic variable ($\Omega$), in $\hat{\Lambda}$, which manifests itself as a weight term when calculating stochastic averages. To be more specific using the gauge-P representation~\cite{gaugeP}, the Ito SDEs are altered such that:
\bea
\label{eq: gauge al}
d \al_{i} & = &   \left( A_{i}^{+}  - g_{k}B_{jk}  \right) dW_{k}  \\
\label{eq: gauge Om}
d \Omega  & = & \Omega \left( V dt + g_{k} dW_{k} \right)
\eea
where summation over $k$ is implied and $V$ is the constant term that may appear after substituting the correspondence relations into an equation of motion for $\hat{\rho}$.

 The gauge-P representation has been very successful in simulating the dynamics of many-mode bose gases~\cite{canonical, finitetemp, correlations} partly because such systems result in neat diagonal noise matrices that are easier to handle as seen in Eq.~\ref{eq: gauge al}. However, it is evidently not as straightforward to apply it in our case as we have a much more complicated non-diagonal noise matrix. The true complication arises when we attempt to calculate Stratonovich correction terms as it is the Stratanovich version of the SDEs that are simulated. We believe that the application of the gauge-P is possible in principle but requires a bit more thought for Heisenberg systems if using the Schwinger boson approach.

\subsection{Finite size effects}
\label{sn: finite size effects}
\begin{figure}[t]
\begin{center}
    \includegraphics{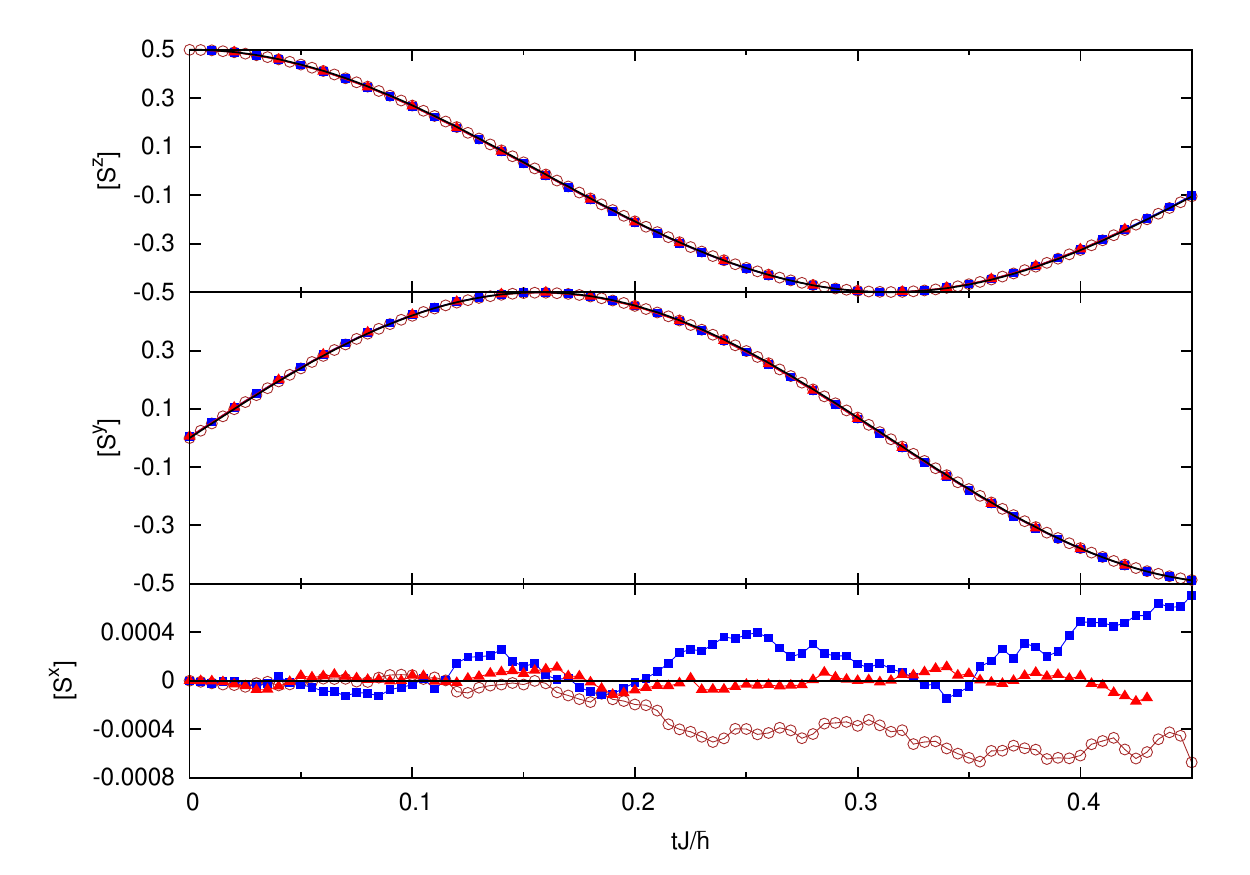}
\end{center}\caption{Isotropic Heisenberg model following a transverse-field quench at $tJ/\hbar=0$ from $h/J=0.0$ to $h/J=10.0$, beginning in the FM ground state: $| \uparrow \uparrow \dots  \uparrow \ra$. FM interactions assumed:  $sign(J)  = +1$.  From top to bottom: plots of $\Sxi, \Syi, \Szi$ vs $tJ/\hbar$ respectively. The stochastic averages, $\la \la . \ra\ra$ are for $N=4$: (\lx), $N=10$: (\lz), $N=100$: (\ly),   while exact diagonlization results for $N=4$ are represented by the black solid line. Simulation parameters: $n_{traj}^{(N=4)} =10^{5}, n_{traj}^{(N=10)} =2\times10^{5}, n_{traj}^{(N=100)} =5\times10^{4},  dt=0.001, \Delta/J=1.0$. Agreement remains good and finite size effects are negligible. The simulations diverge at approximately $tJ/\hbar \sim 0.45$.  }\label{fig: fig5}
\end{figure}

The main advantage of the PPR is the linear scaling with the number of spins, $N$, as compared to the exponentially large matrices needed for an exact solution. We first demonstrate the capabilities of the PPR at simulating large system sizes by showing results for the FM isotropic Heisenberg case at a field value of $h/J=10$, prepared in the initial FM state as in Fig.~\ref{fig: fig4}. As expected, we do not observe any finite size effects within the life time of the stochastic simulations. Even with a chain consisting of $100$ spins (Fig.~\ref{fig: fig5}), the dynamics exhibited are similar to that of a four-spin chain. The simulations are compared against the exact diagonalization results for an $N=4$ system and exhibit identical real time evolution of the spin components for a $1D$ chain with FM interactions, i.e. finite size effects are negligible.

\begin{figure}[h!]
\begin{center}
    \includegraphics{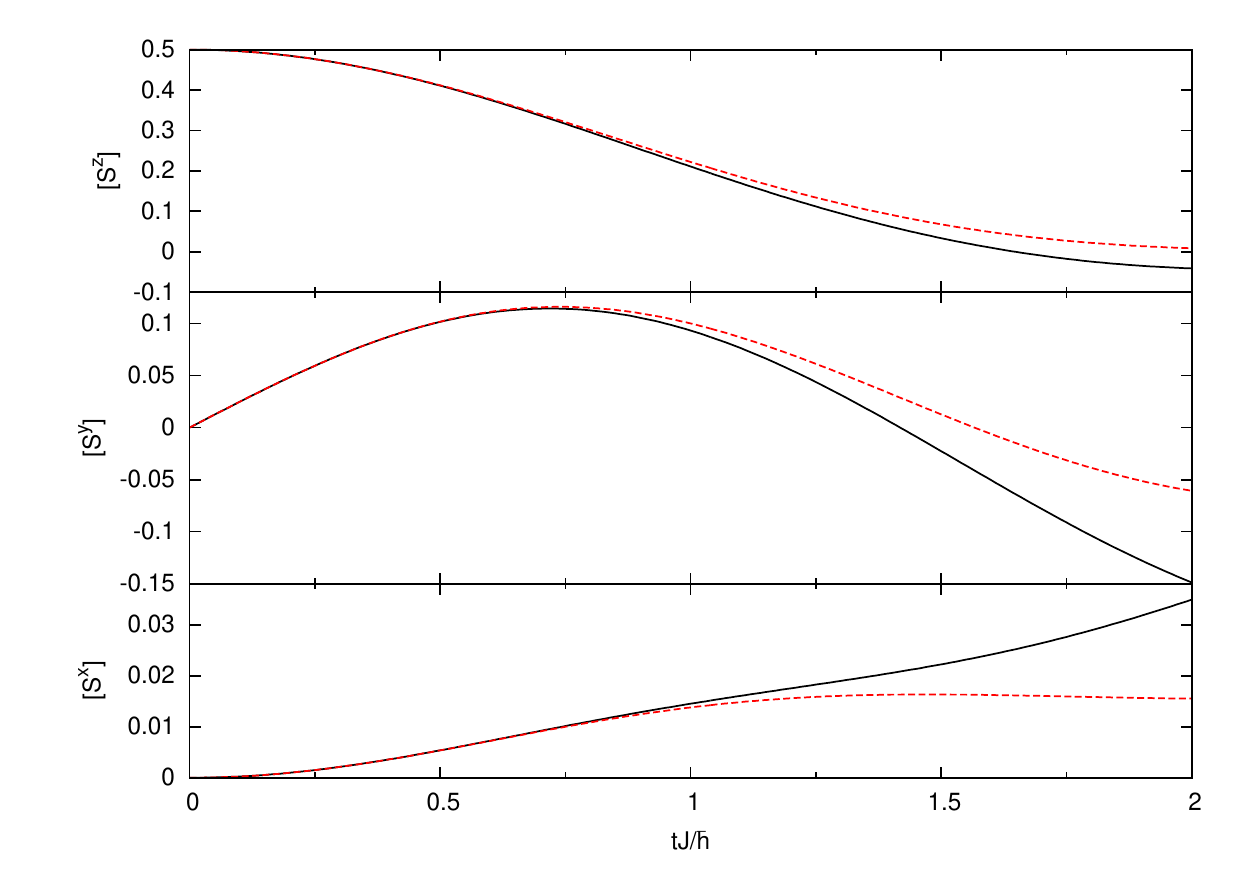}
\end{center}\caption{Anisotropic Heisenberg model following a transverse-field quench at $tJ/\hbar=0$ from $h/J=0.0$ to $h/J=10.0$, beginning in the AFM Neel state: $| \uparrow \downarrow \dots  \uparrow \downarrow \ra$. AFM interactions assumed:  $sign(J)  = -1$.  From top to bottom: plots of $\Sxi, \Syi, \Szi$ vs $tJ/\hbar$ respectively. The exact calculations for the $N=4$ (solid black lines) and $N=10$ (dashed red lines) are compared. We observe $t_{finite}J/ \hbar\sim$0.8 for $\Delta /J \sim -0.8$}\label{fig: fig6}
\end{figure}

\begin{figure}[h!]
\begin{center}
 \includegraphics{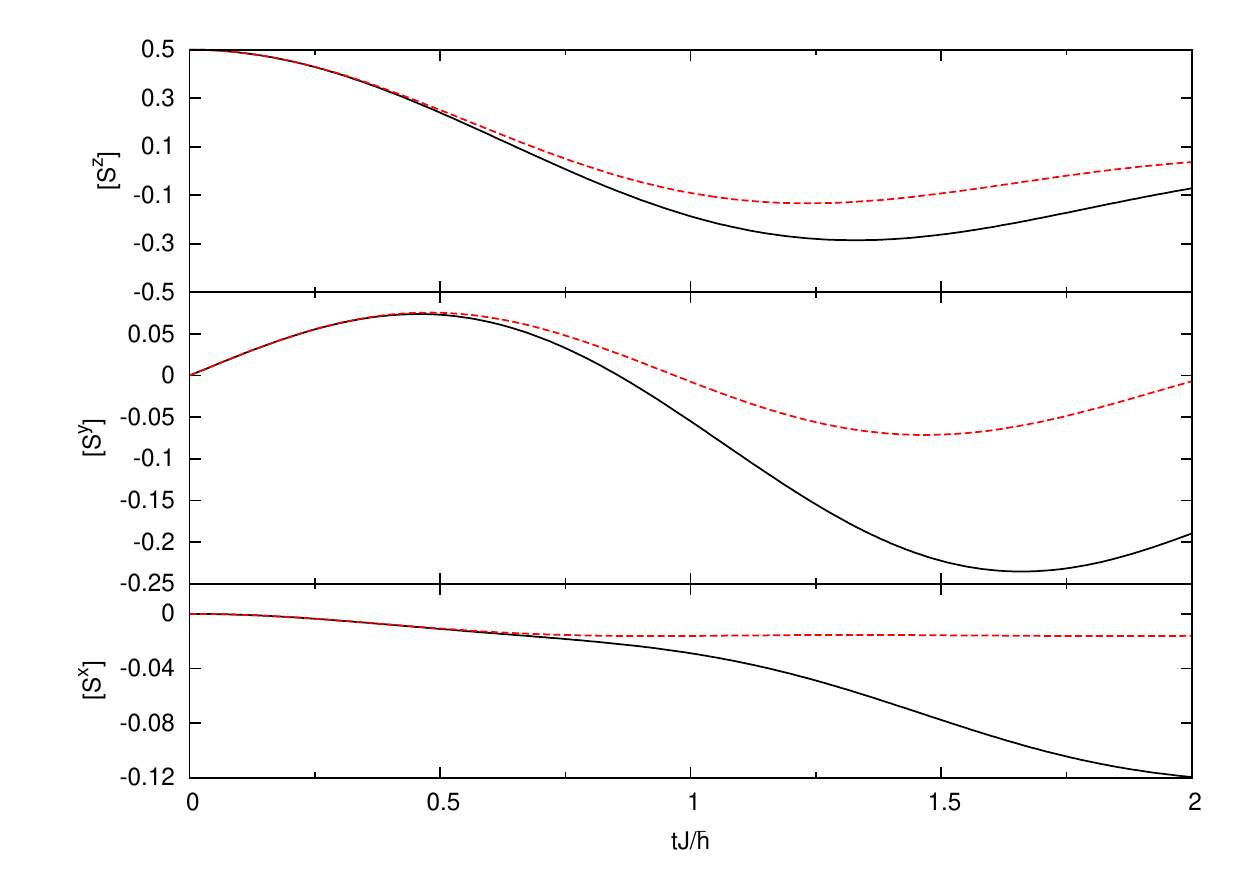}
\end{center}\caption{Anisotropic Heisenberg model following a transverse-field quench at $tJ/\hbar=0$ from $h/J=0.0$ to $h/J=10.0$, beginning in the AFM Neel state: $| \uparrow \downarrow \dots  \uparrow \downarrow \ra$. AFM interactions assumed:  $sign(J)  = -1$.  From top to bottom: plots of $\Sxi, \Syi, \Szi$ vs $tJ/\hbar$ respectively. The exact calculations for the $N=4$ (solid black lines) and $N=10$ (dashed red lines) are compared. We observe $t_{finite}J/ \hbar\sim$ 0.5  for a given anisotropy of $\Delta/J \sim -1.5$} \label{fig: fig7}
\end{figure}

This is not the case for a $1D$ AFM ($J<0$) however. A quantum quench in this model with $h=0$ starting from the N\'eel state has previously been extensively studied~\cite{Barmettler09}. In order to verify that in our approach finite size effects do exist, we performed  $N=4$ and $N=10$ exact calculations  for the anisotropic AFM  with different values of anisotropy: $\Delta/J$. Two sample exact calculations are shown in Fig.~\ref{fig: fig6} and Fig.~\ref{fig: fig7} respectively for $\Delta/J = -0.8 $ and $\Delta/J =-1.5$ for low field values of $h/J=0.5$. For the AFM Heisenberg Hamiltonian, the system is initialized in the classical AFM Neel state: $| \uparrow \downarrow \dots \uparrow \downarrow \ra$. 

An immediate observation is that increasing the value of $\Delta $, reduces the time: $t_{finite}$, which we define as the time that significant finite size effects are noticeable. A natural progression to make in order to take advantage of the SDES we have derived, is to increase the value $\Delta/J$ till $t_{finite} < t_{life}$, thereby allowing us to explore the finite size effects of macroscopically large systems.


\begin{figure}
\begin{center}
    \includegraphics{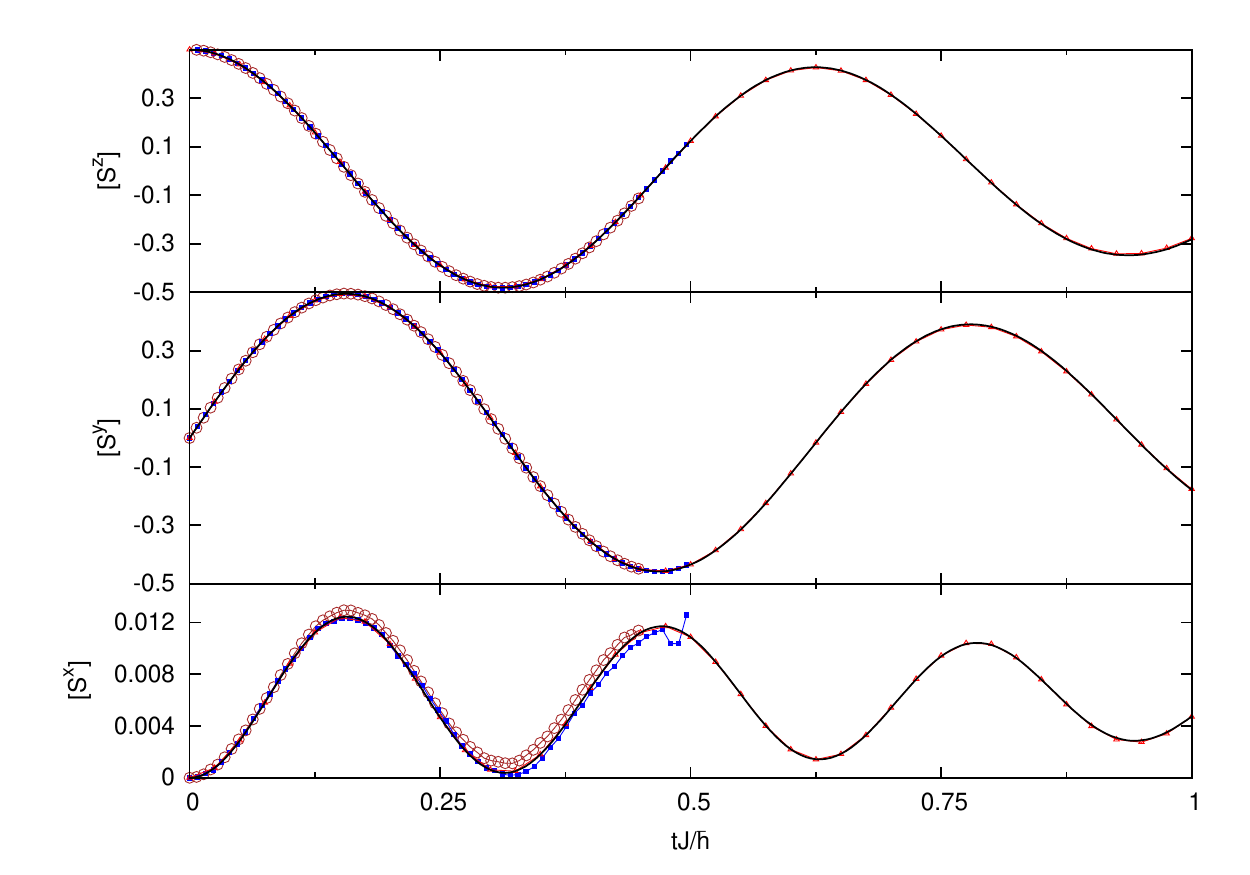}
\end{center}\caption{Anisotropic Heisenberg model following a transverse-field quench at $tJ/\hbar=0$ from $h/J=0.0$ to $h/J=10.0$, beginning in the AFM ground state: $| \uparrow \downarrow \dots  \uparrow \downarrow \ra$. AFM interactions assumed:  $sign(J) =-1$.  From top to bottom: plots of $\Sxi, \Syi, \Szi$ vs $tJ/\hbar$ respectively. The stochastic averages, $\la \la . \ra\ra$ are for $N=4$:( \lx )and $N=10$:( \lz ),   while exact diagonalization results are for $N=4$: (black solid lines) and $N=10$: (\ly). Simulation parameters: $n_{traj}^{(N=4)} =10^{6}, n_{traj}^{(N=10)} =10^{5},  dt=0.001, \Delta/J=-0.5$. Agreement remains good and finite size effects are  unnoticeable up to $tJ\hbar = 1$. The SDEs diverge at $t_{life}J/\hbar \sim 0.48$.}\label{fig: fig8}
\end{figure}

\begin{figure}
\begin{center}
    \includegraphics{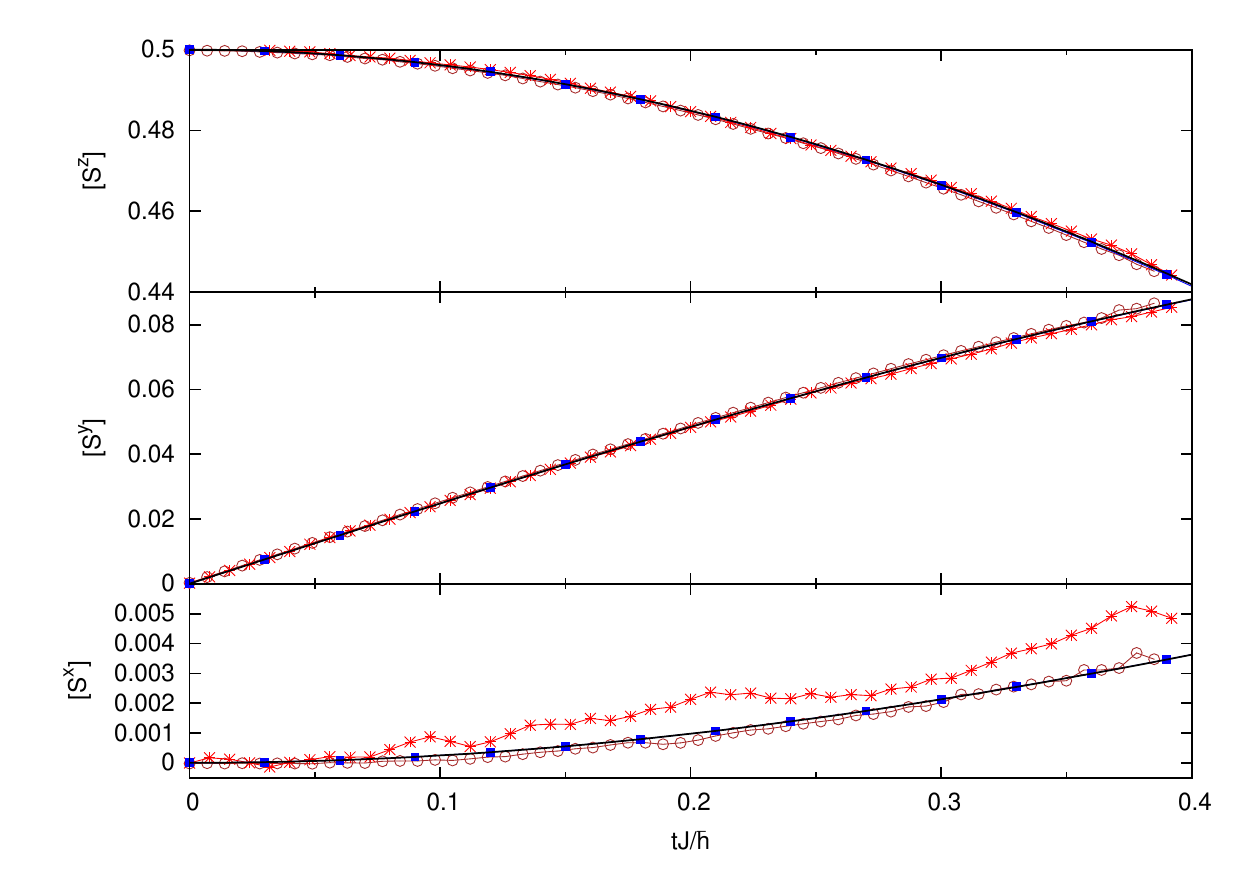}
\end{center}\caption{Anisotropic Heisenberg model following a transverse-field quench at $tJ/\hbar=0$ from $h/J=0.0$ to $h/J=0.5$, beginning in the AFM ground state: $| \uparrow \downarrow \dots  \uparrow \downarrow \ra$. AFM interactions assumed:  $sign(J)  = -1$.  From top to bottom: plots of $\Sxi, \Syi, \Szi$ vs $tJ/\hbar$ respectively. The stochastic averages, $\la \la . \ra\ra$ are for $N=4$:(\lz)and $N=10$:(\lp),   while exact diagonalization results are for $N=4$: (black solid lines) and $N=10$:(\lx). Simulation parameters: $n_{traj}^{(N=4)} =10^{6}, n_{traj}^{(N=10)} =10^{5},  dt=0.001, \Delta/J=-0.8$. Finite size effects are unnoticeable at $t_{life}J/\hbar \sim 0.4$.}\label{fig: fig9}
\end{figure}

We observe finite size effects through the same observables as in eq.~\ref{eq: obs Sx} to eq.~\ref{eq: obs Sz}. However for the initial N\'{e}el state,  it is more meaningful to take into consideration the alternating sign of spins when calculating the averaged spin components\footnote{Note that there is exists an exception. There is no need to account for a sign change for the observable: $\left[ {S}_{x} \right]$}, i.e.:
\be
\Sxi  =  \sum_{i=0}^{N-1} \la \f{1}{2} ( \hat{a}_{i}^{\dg} \hat{b}_{i} + \hat{b}_{i}^{\dg} \hat{a}_{i} )\ra =  \sum_{i=0}^{N-1} \la \la \f{1}{2} (\al^{+}_{i} \beta_{i} + \beta^{+}_{i}\al_{i} ) \ra \ra,
\ee
\be
\Syi  =  \sum_{i=0}^{N-1} \la \f{1}{2i} (-1)^{i} ( \hat{a}_{i}^{\dg} \hat{b}_{i} - \hat{b}_{i}^{\dg} \hat{a}_{i} )\ra =  \sum_{i=0}^{N-1} (-1)^{i} \la \la \f{1}{2i} (\al^{+}_{i} \beta_{i} - \beta^{+}_{i}\al_{i} ) \ra \ra,
\ee
and
\be
\Szi =  \sum_{i=0}^{N-1} \la \f{1}{2} (-1)^{i} ( \hat{a}_{i}^{\dg} \hat{a}_{i} - \hat{b}_{i}^{\dg} \hat{b}_{i} )\ra =  \sum_{i=0}^{N-1}  (-1)^{i} \la \la \f{1}{2} (\al^{+}_{i} \al_{i} - \beta^{+}_{i}\beta_{i} ) \ra \ra.
\ee
Increasing $\Delta / J$ however has the adverse effect of decreasing $t_{life}$ significantly. Thus while it is possible to simulate macroscopically large system sizes, we find that the SDE simulations diverge much sooner than $t_{finite}$. Fig.~\ref{fig: fig8} ($\Delta/J = -0.5, h/J=10.0$) reinforces our claim that $t_{finite}$ decreases with $\Delta/J$ as no finite size effects are observed up to $tJ / \hbar = 1$, in sharp comparison to Fig.~\ref{fig: fig6} ($\Delta/J = -0.8$) and Fig.~\ref{fig: fig7} ($\Delta/J = -1.5$), albeit for $h/J=0.5$. Our last effort to observe finite size effects was to increase $\Delta/J$ to $-0.8$ with hopes that $t_{life} > t_{finite}$. As seen in Fig.~\ref{fig: fig9}, our simulations do not survive beyond $t_{finite}$. Since $t_{finite}$ depends on the anisotropy $\Delta/J$, increasing the system size while possible will result in $t_{life}$ of the same order. In general, we find that increasing $\Delta/J$ will {\it decrease}  $t_{life}$ as well as $t_{finite}$ such that $t_{life} < t_{finite}$ always holds true. This thwarts our efforts on examining finite size effects for the AFM case. Furthermore, we find that using an initial Neel state results in poor convergence for the observable: $\Sxi$ as seen in Fig.~\ref{fig: fig8} (and even more so in Fig.~\ref{fig: fig9}) compared to an initial FM ground state and it is likely that we have used an insufficient number of trajectories in our simulations. Nevertheless, we have demonstrated the applicability of the PPR to AFM systems.

\subsection{Nearest neighbor correlation functions}
\label{ssn: corr functions}
\begin{figure}
\begin{center}
    \includegraphics{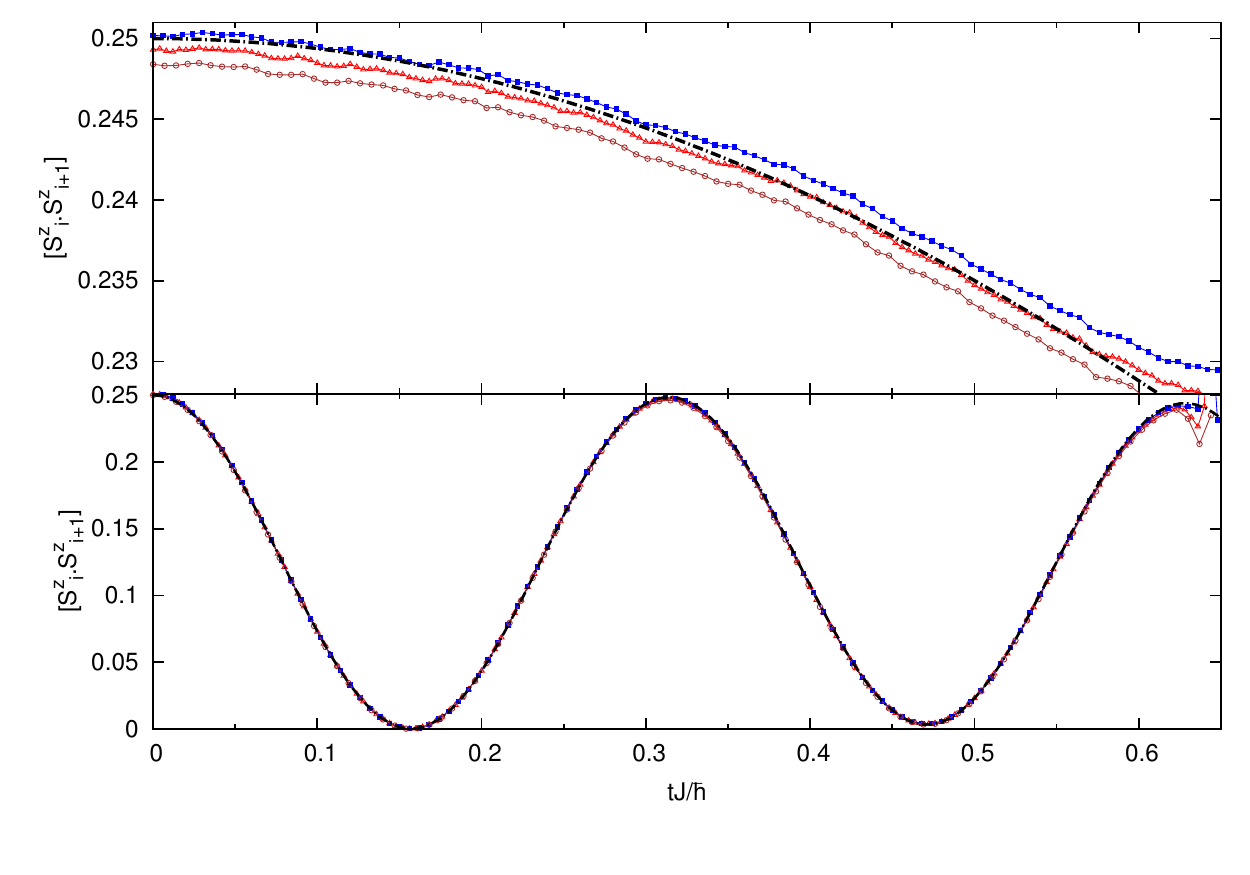}
\end{center}\caption{TFIM following a transverse-field quench at $tJ/\hbar=0$ from $h/J=0.0$ to $h/J=0.5$ (top) and $h/J=0.0$ to $h/J=10.0$ (bottom) beginning in the FM ground state: $| \uparrow \uparrow \dots  \uparrow \uparrow \ra$. FM interactions assumed:  $sign(J)  = 1$.  Plots of $\left[ \hat{S}^{z}_{i} \hat{S}^{z}_{i+1} \right]$ vs $tJ/\hbar$ respectively. The stochastic averages,  $\la \la . \ra\ra$ for $N=4$ are represented by soild red lines (\ly) while the averages plus and minus one standard deviation are presented by (\lx)  and (\lz) lines respectively. Exact diagonalization results are given by dot-dashed black curves. Simulation parameters: $n_{traj} =10^{6} dt=0.001, \Delta/J=0.0$. }\label{fig: fig10}
\end{figure}

Correlation functions are generally of greater interest seeing as they are experimentally accessible quantities. In order to demonstrate the applicability of the PPR in this respect, we calculate the nearest neighbour spin correlation functions for the z-component, which is defined as:
\be
\left[ \hat{S}^{z}_{i} \hat{S}^{z}_{i+1} \right] =  \sum_{i=0}^{N-1}\f { \la  \hat{S}^{z}_{i} \hat{S}^{z}_{i+1}\ra} {N}= \f{1}{4} \sum_{i=0}^{N-1} \f { \la \la \left( n^{\al}_{i} - n^{\beta}_{i}  \right)  \left( n^{\al}_{i+1} - n_{i+1}^{\beta} \right) \ra \ra} {N} ,
\ee
where periodic boundary conditions apply and as before, the following shorthand has been used: $n^{\al}_{i} = \al^{+}_{i}\al_{i}, n^{\beta}_{i} = \beta^{+}_{i}\beta_{i}$. Our calculations in Fig.~\ref{fig: fig10} compares the stochastic averages of correlation functions with the results from exact diagonalization. It is because the correlation function shows poorer convergence than the spin components that we include errorbars for this calculation. Error bars can be calculated from a simple binning analysis of the trajectories and applying the central limit theorem~\cite{gaugeI}. It is not surprising to find poorer convergence for the correlation functions since they amount to higher order moments of the complex phase space variables. Due to the noise terms in the SDEs, the phase space variables are exponentials of gaussian random numbers which are known to diverge sooner for higher moments.



\section{Conclusion}
\label{sn: Conclusion}
We have shown how the real-time quantum
quench dynamics of spin systems can be  simulated via the use of SDEs. This was
done by writing the Heisenberg spin operators in terms of Schwinger bosons and
deriving a Fokker-Planck Equation using the PPR for the density operator. This
in turns allows us to obtain Ito stochastic differential equations which can be
used to calculate the expectation values of normally ordered bosonic operators.
An attractive feature of this prescription is that the number of SDEs scale
linearly as N and can in principle be used to simulate macroscopic system
sizes. In addition, our method is generalizable to higher dimensions and other
geometries as well and explicit couplings to the environment can be included.

The main drawback of the positive-P representation, however is its notoriously
short life time which prevents us from obtaining useful results beyond a
certain time: $t_{life}$. For the TFIM and the anisotropic Heisenberg model, we
found a bare application of the PPR to have $t_{life}J/\hbar\sim 0.45- 0.65$. We
suspect that this is due to drift instability terms present in the SDEs that
cause trajectories to diverge within this time scale.  

We also attempted to explore finite size effects which were more significant for the anisotropic AFM Hamiltonian beginning in the classical N\'{e}el state. For the FM case, no finite size effects were observed even for a lattice size of $100$ spins within its lifetime. We find that the more negative the anisotropy parameter in the AFM Hamiltonian, the sooner finite size effects are observed, i.e. $t_{finite}$ decreases. However, this has the adverse effect of decreasing $t_{life}$ such that $t_{life} < t_{finite}$ for the simulations that we have carried out. 

Finally, we would like to point that in cases where the underlying Hamiltonian has conserved properties, such as the models addressed in this paper, then it could be advantageous to use projection methods instead~\cite{projectionmethod1, projectionmethod2}. This ensures the use of a more efficient basis set which will lead to improved simulation performances. In particular, there exists the PPR approach uses the SU$(n)$ spin coherent states~\cite{drummond} as a basis set instead. 

 An obvious future direction of our research involves applying the gauge-P representation  in a bid to extend simulation life times and to examine the efficient of the other methods suggested above. Also the study of systems in higher dimensions with and without couplings to the environment would be of considerable interest.

\appendix
\section{The positive-P representation}
\label{sn: positiveP}
 In this section, we will review the positive-P representation~\cite{positive} that has been applied to both quantum optics~\cite{SpontEmission, nonlinDamping, secondharmonic} and exact many-body simulations of bose-gases~\cite{150000, manybody1, manybody2, manybody3} successfully. The PPR is already well established and our aim for including this review is simply to provide a self-contained paper for readers who are not as familiar with it.

 In short, the PPR is an expansion of the density operator in terms of an off-diagonal coherent state basis:

\be
\label{eq: positiveP}
\hat{\rho}   =   \int P(\al, \al^{+}) \hat{\Lambda}(\al, \al^{+}) d^{2} \al d^{2} \al^{+}  =     \int P(\al, \al^{+}) \f{| \al \ra \la \al^{+\ast} |}{\la \al^{+\ast}| \al\ra} d^{2} \al d^{2} \al^{+}
\ee
where $|\al \ra = e^{-\f{1}{2} |\al|^{2}} \sum_{n=0}^{\infty} \f{\al^{n}}{n!} |n \ra $ is the standard bosonic coherent state~\cite{glauber} that are eigenstates of the annihilation operator $\hat{a}$. $P(\al, \al^{+})$ plays the role of a distribution function in the phase space spanned by $\left\{\al,\al^{+}\right\}$ and can be chosen such that it remains real and positive. In addition, due to the normalization factor in the denominator of Eq.~\ref{eq: positiveP} and using the fact that $Tr[\hat{\rho}] = 1$, we see that

\be
\int P(\al, \al^{+}) d^{2} \al d^{2} \al^{+} =  1
\ee
i.e. the distribution is normalized over the entire complex phase space. Simply put, we can interpret $P(\al, \al^{+})$ as a probability distribution function for the variables $\al$ and $\al^{+}$, hence the name positive-P.

 A hallmark of  the PPR is that the off-diagonal projection operators, $\hat{\Lambda}(\al, \al^{+})$ satisfies the following correspondence relations:

\bea
\label{eq: correspondence relations}
\hat{a} \hat{\Lambda} &= &\al \hat{\Lambda} \nn \\
\hat{a}^{\dg} \hat{\Lambda} &=&  (\al^{+} + \f{\ptl}{\ptl \al} ) \hat{\Lambda} \\
\hat{\Lambda} \hat{a}^{\dg} &= &\al^{+} \hat{\Lambda} \nn \\
\hat{\Lambda} \hat{a} &=&  ( \al + \f{\ptl}{\ptl \al^{+}}) \hat{\Lambda}. \nn
\eea
which allows us to map complicated operator equations consisting of bosonic annihilation and creation operators onto differential equations of phase space variables $\al, \al^{+}$. The correspondence relation is typically used in an equation of motion for $\hat{\rho}$,  which after integration by parts and ignoring of boundary terms, allows us to obtain a Fokker-Planck equation (FPE):

 \be
 \label{eq: FPE}
 \f{\ptl P(\vec{x})}{\ptl t} = \left\{  -\f{\ptl}{\ptl x^{\mu} }A^{\mu}(\vec{x})  + \f{1}{2} \f{\ptl}{\ptl x^{\mu}} \f{\ptl}{\ptl x^{\nu}} D^{\mu\nu} (\vec{x}) \right\} P(\vec{x}),\ \ , \mu,\nu = 0 \dots N-1,
 \ee
 where $\vec{x}  = \left\{\vec{\al}, \vec{\al}^{+} \right\}$, $A^{\mu}$ is called the drift vector and $D^{\mu \nu}$ is called the diffusion matrix (which is symmetric and positive semi-definite by definition). Due to the doubling of phase space, the diffusion matrix is guaranteed to be positive semi-definite~\cite{positive}. This then allows one to convert the FPE to a set of Ito SDEs proportional to the number of bosonic modes of the system, i.e.

 \be
 \label{eq: Ito SDES}
d {x}^{\mu} = A^{\mu} dt + B^{\mu\nu} dW^{\nu},\ \ \mu= 0 \dots N-1, \nu=0 \dots N_{w},
 \ee
where $dW^{\nu}$ is a vector of Wiener increments with $N_{w}$ components and $B^{\mu \nu}$ is a noise matrix that must satisfy the factorization

\be
\label{eq: factorization}
\mathbf{D} = \mathbf{B} \mathbf{B}^{T}.
\ee
 This factorization is not unique and any noise matrix that satisfies Eq.~\ref{eq: factorization} will produce the same stochastic averages in the limit of an infinite number of trajectories. This ambiguity in the choice of $\mathbf{B}$ may affect the performance of stochastic simulations~\cite{optimization, thesis}.

 Since $\mathbf{D}=\mathbf{D}^{T}$, an obvious factorization to use would be the square root of the diffusion matrix, i.e. $\mathbf{B}= \sqrt{\mathbf{D}}$, which is easily accomplished by using common mathematical software such as Matlab or Maple. While this is the most convenient procedure, it does not necessarily produce the most elegant noise matrix. On the other hand, it is possible to decompose a single diffusion matrix into different diffusion processes~\cite{thesis}: $\mathbf{D} =  \mathbf{D_{1}} + \mathbf{D_{2}} + \mathbf{D_{3}} + \dots$ that may be more easily factorized, i.e. the  factorization $\mathbf{D}_{i} = \mathbf{B_{i}} \mathbf{B_{i}^{T}}$ is trivial. Using this procedure, an equivalent noise matrix that also results in $\mathbf{D}$ is given by

\be
\label{eq: B = [ B1 B2 B3...]}
\mathbf{B} = \left[ \mathbf{B_{1}} \ \mathbf{B_{2}} \ \mathbf{B_{3}} \dots \right].
\ee
Despite possibly taking on a more elegant form, Eq.~\ref{eq: B = [ B1 B2 B3...]} introduces $N_{w}( > N )$ wiener increments and with that the possibility of larger sampling errors. So we see that there are advantages and disadvantages of the two factorization methods.

The convenience in using the positive-P representation is in calculating the expectation values of normal-ordered operators as they can be replaced by simple stochastic averages over their corresponding phase space functions. The equivalence is as follows
\be
\la (\hat{a})^{\dg})^{m} ( \hat{a})^{n}\ra  = \la \la (\al^{+})^{m} (\al)^{n} \ra \ra
\ee
where $\la . \ra$ is the usual quantum mechanical expectation value and $\la\la . \ra\ra$ represents an average over stochastic trajectories. In the limit that the number of trajectories goes to infinity, we get an exact correspondence, although an average over $10^{4}-10^{6}$ trajectories usually gives good agreement\footnote{This is just a general observation of the number of trajectories used in different articles when applying the positive-P representation. See~\cite{nonlinDamping, SpontEmission, 150000} for example. } before sampling errors cause divergences~\cite{gilchrist}.

 The main downside  of the PPR is its notoriously short simulation life times. This is typically caused by instabilities in the drift or  diffusion term~\cite{gaugeP} that cause trajectories to diverge in a finite time,  when a finite number of trajectories are used to calculate expectation values. With that being said, the positive-P representation is best used for systems where the interesting physics occur at short timescales. Nonetheless, this does not deter us from our our current aim of demonstrating the possibility of simulating real time spin dynamics using SDES, even if only for short times.

\section{Initial distribution}
\label{sn:Initial distribution}
 An important point in simulating SDES would be using the right initial values for the phase space variables, $\al, \al^{+}$. For any density matrix, a particular form of the positive-P distribution function~\cite{positive} that always exist is given by
\be
\label{eq: PPR inital distribution}
P(\al, \al^{+})  =  \f{1}{4 \pi^{2}}  \la (\al+ (\al^{+})^{\ast} ) /2 | \hat{\rho} | ( \al + ( \al^{+})^{\ast}  )/2 \ra   e^ { - |\al - ( \al^{+})^{\ast}  |^{2}/4}.
\ee
 It has been shown in~\cite{numberstates} that using Eq.~\ref{eq: PPR inital distribution}, it is possible to initialize the phase space variables for a variety of initial states such as: coherent states, fock states or crescent states to name a few. Of interest to us is the initial positive P-distribution for number states: $|n\ra\la n |$ which takes the form:
\be
\label{eq: fockP}
P(\mu, \gm)= \f{e^{-|\gm|^{2} } } {\pi} \f{\Gm(|\mu|^{2}, n+1  )} {\pi}
\ee
where
\be
\Gm(x,n) = \f{e^{-x} x^{n-1}}{(n-1)!}
\ee
is the Gamma distribution.  Our phase space variables are related to $\gm$ and $\mu$ via the relation $\al = \mu + \gm$ and $\al^{+} = \mu^{\ast} - \gm^{\ast}$ and so by sampling $\gm$ and $\mu$ using the appropriate distribution functions in Eq.~\ref{eq: fockP} (i.e. gamma distribution for $\mu$ and gaussian distribution for $\gm$), we can invert them to find the numerical values for $\al$ and $\al^{+}$ that represents the fock state $|n\ra \la n |$. Although, we have only outlined the steps for initializing the distribution of a fock state,  more explicit details can be found in the useful article in~\cite{numberstates} . 

While in this paper, we initialise the system in either the FM ground state or the AFM, it is in principle possible to initialise the system in a general entangled state, which is described by the following density operator:
\be
\label{eq: bell state}
\hat{\rho} = \f{1}{N} \left(  w_{1} |\uparrow \ra +  w_{0}| \downarrow \ra \right) \left( \la\uparrow|  w_{1} + \la  \downarrow| w_{0} \right), 
\ee
where $N = w_{0}^{2} + w_{1}^{2}$ and $w_{0}$ and $w_{1}$ represent the probabities of the entangled state being spin down and spin up state respectively. Or in the language of $\hat{a}$ and $\hat{b}$ bosons:
\be
\label{eq: entangled state}
\hat{\rho} = \f{1}{N} \left( w_{1}|1,0 \ra +  w_{0}| 0,1 \ra \right) \left( \la 1,0| w_{1}  + \la  0,1| w_{0} \right).
\ee
The general entangled state is of interest as it is the ground state of the random field Ising model (RFIM), which our formalism is also able to address. Substituting eq.~\ref{eq: entangled state} into eq.~\ref{eq: PPR inital distribution}, the coherent state basis results in the following expression for the probability distribution:
\bea
\fl P(\mu_{\vec{\al}}, \gamma_{\vec{\al}}, \mu_{\vec{\beta}}, \gamma_{\vec{\beta}} ) & = & \f{1}{N} \left[ w_{0}^{2} \Gamma(|\mu_{\vec{\al}}|^{2},2) \f{e^{-|\gamma_{\vec{\al}}|^{2}}}{\pi}
 \delta(\mu_{\vec{\beta}}) +  w_{1}^{2} \delta(\mu_{\vec{\al}}) \Gamma(|\mu_{\vec{\beta}}|^{2},2 ) \f{e^{-|\gamma_{\vec{\beta}}|^{2}}}{\pi}\right. \nn \\
& & \left. w_{0}w_{1} \left( \f{ e^{-|\mu_{\al}|^{2}} |\mu_{\vec{\al}}| } {\pi} \f{e^{-|\gamma_{\vec{\al}}|^{2}}}{\pi} \right) \left( \f{e^{-|\mu_{\beta}|^{2}} |\mu_{\vec{\beta}}|}{\pi} \f{e^{-|\gamma_{\vec{\beta}}|^{2}}}{\pi} \right) 2\cos(2 \eta) \right]
\eea	
where $\mu_{\vec{\al}} = |\mu_{\vec{\al}}|e^{i(\eta+ \xi)} = |\mu_{\vec{\al}}|e^{i(\xi+ \eta)}$ and $\mu_{\vec{\beta}} = |\mu_{\vec{\beta}}|e^{i(\eta+ \xi)} = |\mu_{\vec{\beta}}|e^{i(\xi -\eta)}$. Note that we have made a similar change of variables as above, i.e. 
\bea
\mu_{\vec{\al}}  =   \f{ \al +(\al^{+})^{\ast} }{2}, & & \  \ \gamma_{\vec{\al}}  =  \f{ \al -(\al^{+})^{\ast} }{2}  \nn \\
\mu_{\vec{\beta}}  =   \f{ \beta +(\beta^{+})^{\ast} }{2}, & & \ \ \gamma_{\vec{\beta}}  =  \f{ \beta-(\beta^{+})^{\ast} }{2} 
\eea
The PPR based on the SU-(n) coherent states~\cite{drummond} seems more tailored to dealing with superposition states, as they can be more easily initialized with delta functions.

\section{Fokker-Planck Equation for Heisenberg Hamiltonian}
\label{sn: additional details}
 If we were to apply formalism outlined in~\ref{sn: positiveP}, we obtain the following FPE for the TFIM in Eq.~\ref{eq: TFIM}:

\bea
\label{eq: full FPE}
\fl \f{\ptl P(\vec{\al}, \vec{\al}^{+}, \vec{\beta}, \vec{\beta}^{+})}{dt} & = & \sum_{i} \left( -\f{\ptl}{\ptl \al_{i}}\left\{ \f{i J}{4\hbar} \al_{i} \left[ (n_{i+1}^{\al} - n_{i+1}^{\beta}) + (n_{i-1}^{\al} - n^{\beta}_{i-1} ) \right] +\f{i h(t)}{2\hbar} \beta_{i}\right\} \right.\nn  \\
& &  - \f{\ptl}{\ptl \al_{i}^{+}}\left\{  \f{i J}{4\hbar} \al^{+}_{i} \left[ (n_{i+1}^{\beta} - n^{\al}_{i+1}) + (n_{i-1}^{\beta} - n^{\al}_{i-1} )  \right]-\f{i h(t)}{2\hbar} \beta_{i}^{+} \right\} \nn \\
& & - \f{\ptl}{\ptl \beta_{i}}\left\{ \f{i J}{4\hbar}  \beta_{i} \left[ (n_{i+1}^{\beta} - n^{\al}_{i+1}) + (n_{i-1}^{\beta} - n^{\al}_{i-1} ) \right] +\f{i h(t)}{2\hbar} \al_{i} \right\} \nn \\
& & -\f{\ptl}{\ptl \beta^{+}_{i}}\left\{ \f{i J}{4\hbar}\beta_{i}^{+} \left[ (n_{i+1}^{\al} - n^{\beta}_{i+1}) + (n_{i-1}^{\al} - n^{\beta}_{i-1} ) \right] -\f{i h(t)}{2\hbar} \al_{i}^{+} \right\} \nn \\
\fl & \fl & \fl + \f{1}{2} \left( \f{iJ}{4 \hbar} \right) \left[ \f{\ptl^{2}} {\ptl \al_{i} \ptl \al_{i+1}}\al_{i} \al_{i+1}   + \f{\ptl^{2}} {\ptl \al_{i+1}\ptl \al_{i} }\al_{i} \al_{i+1}  - \f{\ptl^{2}} {\ptl \al^{+}_{i} \ptl \al^{+}_{i+1}}\al^{+}_{i} \al^{+}_{i+1}   - \f{\ptl^{2}} {\ptl \al^{+}_{i+1}\ptl \al^{+}_{i} }\al^{+}_{i} \al^{+}_{i+1}  \right. \nn \\
& & \fl \left.\f{\ptl^{2}} {\ptl \beta_{i} \ptl \beta_{i+1}}\beta_{i} \beta_{i+1}   + \f{\ptl^{2}} {\ptl \beta_{i+1}\ptl \beta_{i} }\beta_{i} \beta_{i+1}  - \f{\ptl^{2}} {\ptl \beta^{+}_{i} \ptl \beta^{+}_{i+1}}\beta^{+}_{i} \beta^{+}_{i+1}   - \f{\ptl^{2}} {\ptl \beta^{+}_{i+1}\ptl \beta^{+}_{i} }\beta^{+}_{i} \beta^{+}_{i+1}  \right. \nn \\
& &\fl \f{\ptl^{2}} {\ptl \al^{+}_{i} \ptl \beta^{+}_{i+1}}\al^{+}_{i} \beta^{+}_{i+1}   + \f{\ptl^{2}} {\ptl \beta_{i+1}^{+}\ptl \al^{+}_{i} } \al^{+}_{i} \beta^{+}_{i+1}  + \f{\ptl^{2}} {\ptl \al^{+}_{i+1} \ptl \beta^{+}_{i}} \al^{+}_{i+1} \beta^{+}_{i}   + \f{\ptl^{2}} {\ptl \beta_{i}^{+}\ptl \al^{+}_{i+1} }\al_{i+1}^{+} \beta_{i}^{+} \nn \\
&  & \fl \left.  \left.- \f{\ptl^{2}} {\ptl \al_{i} \ptl \beta_{i+1}}\al_{i} \beta_{i+1}   -  \f{\ptl^{2}} {\ptl \beta_{i+1}\ptl \al_{i} }\al_{i} \beta_{i+1} -  \f{\ptl^{2}} {\ptl \al_{i+1} \ptl \beta_{i}}\al_{i+1} \beta_{i}  \right. \right.\nn \\
&  & \fl  \left. \left.  - \f{\ptl^{2}} {\ptl \beta_{i}\ptl \al_{i+1} }\al_{i+1} \beta_{i}  \right] \right) P(\vec{\al}, \vec{\al}^{+}, \vec{\beta}, \vec{\beta}^{+})
\eea
where we have already carried out an integration by parts and assumed that boundary terms vanish. By inspecting Eq.~\ref{eq: full FPE},  the diffusion matrix (which is a $4N \times 4N$ matrix),  has matrix elements that are specified by the functions associated with their derivatives.

 Obviously, calculating the noise matrix is not a trivial task and comprises the bulk of the analytical work.  Instead of simply taking the straightforward $\mathbf{B} = \sqrt{ \mathbf{D}}$ choice, we used the trick mentioned in~\ref{sn: positiveP} and decomposed our diffusion matrix into eight different constituents,  i.e.:
\be
\mathbf{D} =  \mathbf{D^{\al}} +  \mathbf{D^{\beta}} +  \mathbf{D^{\al^{+}}} +  \mathbf{D^{\beta^{+}}} +  \mathbf{D^{ \beta\al}}  + \mathbf{D^{\al \beta}}+ \mathbf{D^{\beta^{+} \al^{+}}} + \mathbf{D^{\al^{+} \beta^{+}}}
\ee
where the obvious choice for these constituents would be
\bea
\left(\mathbf{D^{\al}} \right)_{i, i+1} & = \left(\mathbf{D^{\al}} \right)_{i+1, i} =  &  \f{iJ}{4\hbar}\al_{i} \al_{i+1} \nn \\
\left(\mathbf{D^{\beta}} \right)_{i, i+1} & =  \left(\mathbf{D^{\beta}} \right)_{i+1, i} = &  \f{iJ}{4\hbar}\beta_{i} \beta_{i+1} \nn \\
\left(\mathbf{D^{\al^{+}}} \right)_{i, i+1} & =  \left(\mathbf{D^{\al^{+}}} \right)_{i+1, i} = & -\f{iJ}{4\hbar}  \al^{+}_{i} \al^{+}_{i+1} \nn \\
\left(\mathbf{D^{\beta^{+}}}\right)_{i, i+1} & =  \left(\mathbf{D^{\beta^{+}}} \right)_{i+1, i} = & -\f{iJ}{4\hbar} \beta^{+}_{i} \beta^{+}_{i+1} \nn \\
\left(\mathbf{D^{\beta \al}}\right)_{i, i+1} & =  \left(\mathbf{D^{\beta \al}} \right)_{i+1, i} = & -\f{iJ}{4\hbar} \al_{i} \beta_{i+1} \nn \\
\left(\mathbf{D^{\al \beta}}\right)_{i, i+1} & = \left(\mathbf{D^{\al \beta}} \right)_{i+1, i} =  & -\f{iJ}{4\hbar} \beta_{i} \al_{i+1} \nn \\
\left(\mathbf{D^{\beta^{+} \al^{+}}}\right)_{i, i+1} & = \left(\mathbf{D^{\beta^{+} \al^{+} }} \right)_{i+1, i} =  &  \f{iJ}{4\hbar}\al^{+}_{i} \beta^{+}_{i+1} \nn \\
\left(\mathbf{D^{\al^{+} \beta^{+}}}\right)_{i, i+1} & = \left(\mathbf{D^{\al^{+} \beta^{+} }} \right)_{i+1, i} =  &  \f{iJ}{4\hbar}\beta^{+}_{i} \al^{+}_{i+1}. \nn
\eea
The idea is that instead of factorizing one complicated diffusion matrix, $\mathbf{D}$ we can instead factorize eight relatively simpler looking noise matrices, i.e. solving $\mathbf{B^{x}} \left( \mathbf{B^{x}} \right)^{T} = \mathbf{D^{x}}$. To make things slightly more transparent we will write out the general form for the first constituent, i.e. $x=\al$:
\be
\mathbf{D^{\al}} =\f{iJ}{4\hbar} \left[ \ba{cccc }
\left[  \ba{ccccc} 0   & {\al_{0} \al_{1}} & 0   & \dots  & t{\al_{0} \al_{N-1}}    \\
 {\al_{1} \al_{0}} & 0 &  {\al_{1} \al_{2}} & \dots 		& 0 \\
0 &  {\al_{2} \al_{1}}&  0 & \ddots & 0  \\
\vdots &  0 & \ddots & \dots & 0   \\
{\al_{N-1} \al_{0}} &  0 & \dots & \dots &  0   \ea \right]  &  \mathbf{0}    & \mathbf{0}   & \mathbf{0}  \\
 \mathbf{0}  &  \mathbf{0}    & \mathbf{0}   & \mathbf{0}  \\
 \mathbf{0}  &  \mathbf{0}    & \mathbf{0}   & \mathbf{0}  \\
 \mathbf{0}  &  \mathbf{0}    & \mathbf{0}   & \mathbf{0}  \\
\ea \right]
\ee
where $\mathbf{0}$  represents an $N\times N$ null matrix. If it were possible to find $\mathbf{B}^{x}$ for all $x$, then the total noise matrix takes the form of Eq.~\ref{eq: B = [ B1 B2 B3...]}.

 Unfortunately, using the obvious choice $\sqrt{B^{x}}$ would still be messy and it would appear that we have not made things any easier. However, we can apply the same trick once more and decompose each $\mathbf{D^{x}}$ into $N$ subconstituents: $\left\{ \mathbf{D^{x}_{j}}, \ \ j=0\dots N-1 \right\}$.  Once again taking the $x=\al$ matrix as an example, the intuitive way of choosing the subconstituents is:
\bea
\label{eq: diffusion factorization}
 \mathbf{D^{\al} }  &   = &   \mathbf{D^{\al}_{0} } + \mathbf{D^{\al} _{1}} + \dots +  \mathbf{D^{\al} _{N-1}}  \\ \nn \\
& \fl = &  \fl   \f{iJ}{4\hbar}\left[ \ba{ccccc}
0 & \al_{0} \al_{1} & \dots &  \dots & 0  \\
\al_{1}\al_{0} & 0 & \dots &  \dots &  \\
\vdots  & \vdots  & \ddots & \dots & \vdots  \\
\vdots  & \vdots  & \ddots & \dots & \vdots  \\
0  &  0   & \dots & \dots & 0  \\
\ea\right] +
\f{iJ}{4\hbar}\left[ \ba{ccccc}
0 &  0 & \dots &  \dots & 0  \\
0 & 0 &  \al_{1}\al_{2} &  \dots &  \\
0  &  \al_{2}\al_{1}  & \ddots & \dots &  \\
\vdots  &   \vdots   & \dots & \ddots & \vdots  \\
0 & \dots  & \dots & \dots  & 0
\ea\right]
+ \dots  \\
&  &
+ \f{iJ}{4\hbar} \left[ \ba{ccccc}
0 &  \dots  & \dots &  \dots & { \al_{0} \al_{N-1} }   \\
0 & \dots  &  \vdots &  \dots &  \\
\vdots   &  \dots  & \ddots & \dots &  \\
\vdots  &   \vdots   & \dots & \ddots & \vdots  \\
{\al_{N-1} \al  } & \dots  & \dots & \dots  & 0
\ea\right]
\nn \eea
where the only non-trivial matrix elements of $\mathbf{D^{\al}_{j}}$ are given by
\be
(\mathbf{D^{\al}_{j}})_{i,i+1}   =  (\mathbf{D^{\al}_{j}})_{i+1,i}  = \f{iJ}{4\hbar} \al_{j} \al_{j+1}
\ee
Each subconstituent diffusion matrix $\mathbf{D}^{\al}_{i}$ can then be individually factorized. This reduces the original problem to the much more trivial problem of factorizing matrices of the following form:

\be
\mathbf{D^{\pr}}  = \left[\ba{cc} 0  & X \\ X & 0   \ea \right]
\ee
for which we can easily show that either

\be
\mathbf{B^{\pr}} =  \left[\ba{cc} -\sqrt{X/2}  & -i\sqrt{X/2} \\ -\sqrt{X/2} &   i \sqrt{X/2}   \ea \right]
\ee
or

\be
\mathbf{B^{\pr \pr}} =  \left[\ba{cc} -\sqrt{X/2}  &  i\sqrt{X/2} \\ -\sqrt{X/2} &  - i \sqrt{X/2}   \ea \right]
\ee
satisfies the necessary relation in Eq.~\ref{eq: factorization}. Now, granted that the decomposition for each $\mathbf{D}^{\al}_{i}$ exists, we can write Eq.~\ref{eq: diffusion factorization} as:

\be
\mathbf{D^{\al}} =  \mathbf{B^{\al}_{0}} \left( \mathbf{B^{\al}_{0}} \right)^{T} +   \mathbf{B^{\al}_{1}} \left( \mathbf{B^{\al}_{1}} \right)^{T} + \dots +  \mathbf{B^{\al}_{N-1}} \left( \mathbf{B^{\al}_{N-1}} \right)^{T}
\ee
so that according to Eq.~\ref{eq: B = [ B1 B2 B3...]}, the total noise matrix for $\mathbf{D^{\al}}$ takes the obvious form:
\be
\label{eq: total noise matrix}
\mathbf{B^{\al}} =
\left[
\ba{cccccc}
\vline & \vline & \vline & \vline & \vline& \vline \\
\mathbf{B_{0}^{\al}} & \mathbf{B_{1}^{\al}}  &\dots & \mathbf{B_{j}^{\al}} & \dots & \mathbf{B_{N-1}^{\al}} \\
\vline & \vline & \vline & \vline & \vline& \vline
\ea \right]
\ee
 obviously satisfying Eq.~\ref{eq: factorization}, with the only non-zero elements being:
\bea
\left( \mathbf{B_{j}^{\al}} \right)_{j,2j} & = &  - \f{1}{2}\sqrt{ \f{iJ}{4\hbar} }\sqrt{\al_{j} \al_{j+1}}  \nn \\
\left( \mathbf{B_{j}^{\al}} \right)_{j,2j+1} & = &  - \f{i}{2}\sqrt{ \f{iJ}{4\hbar} }\sqrt{\al_{j} \al_{j+1}}  \nn \\
\left( \mathbf{B_{j}^{\al}} \right)_{j+1,2j} & = &  -\f{1}{2}\sqrt{ \f{iJ}{4\hbar} }\sqrt{\al_{j} \al_{j+1}}  \nn \\
\left( \mathbf{B_{j}^{\al}} \right)_{j+1,2j+1} & = &  \f{i}{2}\sqrt{ \f{iJ}{4\hbar} } \sqrt{\al_{j} \al_{j+1}}  \nn
\eea
where $j=0\dots N-1$. As an explicit example, the $N=4$ case of Eq.~\ref{eq: total noise matrix} is shown below:

\begin{equation}
\fl \mathbf{B^{\al}}  =
\ba{ccc}
\f{1}{2}\sqrt{ \f{iJ}{4\hbar} } \left[ \left[ \ba{cc}
-\sqrt{\al_{0}\al_{1}} &  - i \sqrt{\al_{0}\al_{1}}  \\
 -\sqrt{\al_{0}\al_{1}} &  + i \sqrt{\al_{0}\al_{1}} \\
0 & 0 \\
0 & 0 \\
\vdots & \vdots \\
0 & 0  \ea \right] \right.
&
\left[ \ba{cc}
0 & 0 \\
 -\sqrt{\al_{1}\al_{2}} &  - i \sqrt{\al_{1}\al_{2}} \\
-\sqrt{\al_{1}\al_{2}} &   i \sqrt{\al_{1}\al_{2}}  \\
0 & 0 \\
\vdots & \vdots \\
0 & 0  \ea \right]
&
\left[ \ba{cc}
0 & 0 \\
0 & 0 \\
-\sqrt{\al_{2}\al_{3}} &  - i \sqrt{\al_{2}\al_{3}}  \\
 -\sqrt{\al_{2}\al_{3}} &   i \sqrt{\al_{2}\al_{3}} \\
\vdots & \vdots \\
0 & 0  \ea \right]
\ea
\end{equation}
\be
 \left. \left[ \ba{cc}
-\sqrt{\al_{0}\al_{3}} &  - i \sqrt{\al_{0}\al_{3}}  \\
0 & 0 \\
0 & 0 \\
 -\sqrt{\al_{0}\al_{3}} &  + i \sqrt{\al_{0}\al_{3}} \\
\vdots & \vdots \\
0 & 0 \\
   \ea \right]   \right]
\ee
which is an $4N \times 2N$ matrix with most elements being trivial. This noise matrix would therefore introduce $2N$ independent Wiener increments (see Eq.~\ref{eq: Ito SDES}) can be stored as the components of the Wiener increment vector: $d\vec{W}^{\al}$. In this fashion, the noise terms for the SDEs in Eq.~\ref{eq: SDE al} to Eq.~\ref{eq: SDE beta+} can be derived. If we label $d\vec{W}^{\al}$ in the conventional way\footnote{The labeling for $dW^{\al\beta}, dW^{\beta\al}, dW^{\beta^{+} \al^{+}}$ and $dW^{\al^{+}\beta^{+}}$ does not follow the usual convention and can be deduced from the corresponding noise terms in Eq.~\ref{eq: SDE al} to Eq.~\ref{eq: SDE beta+}} then:

\be
d\vec{W}^{\al} = \left[ \ba{c} dW^{\al}_{0} \\   dW^{\al}_{1} \\ \vdots \\ \vdots \\  dW^{\al}_{N-1}  \ea\right],
\ee
and the resulting stochastic terms only contribute to $d\vec{\al}$, i.e.:

\be
d \al_{i} \propto -\sqrt{\al_{i} \al_{i+1} }\left( dW_{2i} + i dW_{2i+1} \right) - \sqrt{\al_{i} \al_{i-1}} \left( dW_{2i-2} + i dW_{2i-1} \right).
\ee
where we assumed "periodic boundary conditions" for the Wiener increment vectors in the sense that $dW_{-i} = dW_{N-i}$ where $i \in [0,N-1]$.

\ack{This research has been made possible with the support of an NSERC research grant. We would also like to thank  Piotr Deuar and Murray Olsen for helpful discussions.}
\section*{References}
\bibliographystyle{unsrt}
\bibliography{Sbib}

\end{document}